\documentclass[usenatbib]{mnras}

\usepackage[utf8]{inputenc}

\usepackage[T1]{fontenc}
\usepackage{ae,aecompl}

\usepackage[x11names, rgb, table]{xcolor}

\usepackage{pifont} 

\usepackage{graphicx}	
\graphicspath{{./plots/}}
\usepackage{amsmath}	
\usepackage{amssymb}	

\usepackage{tikz} 
\usepackage{pgf} 
\usepackage{booktabs}

\newcommand{\cmark}{\ding{51}}%
\newcommand{\xmark}{\ding{55}}%

\definecolor{chromeyellow}{rgb}{1.0, 0.35, 0.0}
\definecolor{MP}{rgb}{0.18, 0.545, 0.341}
\newcommand{\revision}[1]{#1}

\title[Finding Black Holes with Black Boxes]{Finding Black Holes with Black Boxes -- Using Machine Learning to Identify Globular Clusters with Black Hole Subsystems}

\author[A. Askar et al.]{
Ammar Askar$^{1}$\thanks{E-mail: aaskar@purdue.edu},
Abbas Askar$^{2}$\thanks{E-mail: askar@astro.lu.se},
Mario Pasquato$^{3}$ and
Mirek Giersz$^{4}$ \\
$^{1}$Department of Computer Science, College of Science, Purdue University, 305 N. University Street, West Lafayette, IN 47907, USA\\
$^{2}$Lund Observatory, Department of Astronomy, and Theoretical Physics, Lund University, Box 43, SE-221 00 Lund, Sweden\\
$^{3}$INAF, Osservatorio Astronomico di Padova, vicolo dell’Osservatorio 5, 35122 Padova, Italy \\
$^{4}$Nicolaus Copernicus Astronomical Center, Polish Academy of Sciences, ul. Bartycka 18, 00-716 Warsaw, Poland
}

\date{Accepted XXX. Received YYY; in original form ZZZ}

\pubyear{2018}

\begin{document}
\label{firstpage}
\pagerange{\pageref{firstpage}--\pageref{lastpage}}
\maketitle

\begin{abstract}
Machine learning is a powerful technique, becoming increasingly popular in
astrophysics. 
In this paper, we apply machine learning to more than a thousand globular cluster (GC) models simulated with the \textsc{mocca} project in order to correlate present-day observable properties with the presence
of a subsystem of stellar mass black holes (BHs). 
The machine learning model is then applied to available observed parameters for Galactic GCs to identify which of them that are most likely to be hosting a sizeable number of BHs and 
reveal insights into what properties lead to the formation of
BH subsystems. With our machine learning model, we were able to shortlist 18 Galactic GCs that are most likely to contain a BH subsystem. We show that the clusters shortlisted by the machine learning classifier include those in which BH candidates have been observed (M22, M10 and NGC 3201) and that our results line up well with independent simulations and previous studies that manually compared simulated GC models with observed properties of Galactic GCs. These results can be useful for observers searching for elusive stellar mass BH candidates in GCs and further our understanding of the role BHs play in GC evolution. In addition, we have released an online tool that allows one to get predictions from our model after they input observable properties.

\end{abstract}




\section{Introduction}

The use of machine learning is rapidly increasing in astronomy, with applications including the modeling of instrumental systematics \citep[][]{2012MNRAS.419.2683G} and the
classification of variable-stars with over 24\% improvement over prior methodologies \citep{2011ApJ...733...10R}. Supervised machine learning models in astronomy have mostly been used on hand-labeled data in order to automate tedious tasks on large survey datasets. In this work, we demonstrate a different approach: we train models on realistic simulations and attempt inference on real data. Through our approach, we try to identify Galactic globular clusters (GCs) that are most likely to be harbouring a large number of stellar mass black holes (BHs). By identifying clusters that are likely to contain BHs, we hope to
achieve two main goals. Firstly, our findings will assist observers
searching for stellar mass BHs by narrowing down their search space to a few 
likely candidate clusters. Secondly, we hope to better understand
the dynamical history of clusters containing BHs by identifying and studying them individually.

The motivation for finding these clusters is brought on by the growing observational evidence that dense stellar systems like GCs contain stellar mass BHs. A few accreting BH candidates have been identified through radio and X-ray observations in the Galactic GCs M22 \citep{2012Natur.490...71S}, M62 \citep{2013ApJ...777...69C}, 47 Tuc \citep{2015MNRAS.453.3918M,2017MNRAS.467.2199B} and M10 \citep{2018ApJ...855...55S}. \citet{2018MNRAS.475L..15G} discovered a BH in the Galactic GC NGC 3201 from spectroscopically obtained radial velocity measurements of its main sequence companion star. Accreting BH candidates have also been identified in few extragalactic GCs mainly through X-ray observations \citep{2007Natur.445..183M,2008ApJ...689.1215B,2011MNRAS.410.1655M,2011ApJ...734...79B,2018ApJ...862..108D}. Dynamical mass estimates for some Galactic GCs \citep{2016MNRAS.462.1937S} and several extragalactic GCs \citep{2015ApJ...805...65T} reveal the presence of a significant fraction of non-luminous matter in those GCs. Furthermore, an isolated stellar mass BH candidate in  NGC 6553 was discovered through gravitational microlensing \citep{2015ApJ...810L..20M}. All these observational discoveries point towards the possibility that there could be many more undiscovered BHs in GCs. 

Depending on the initial mass of a GC, up to thousands of BHs should have formed from the evolution of massive stars in these dense stellar systems within tens of millions of years. Whether or not a significant fraction of BHs is retained in GCs depends on the natal kicks which BHs receive and the escape velocity from the GC. Distribution of BH natal kicks is uncertain and weakly constrained
\citep{2009ApJ...697.1057F,2013A&A...556A...9Z,2014ApJ...790..119W,2015A&A...573A..58Z,2016MNRAS.456..578M,2017NewAR..78....1M,2017MNRAS.467..298R,2017PhRvL.119a1101O,2018PhRvD..97d3014W,2017MNRAS.468.4968A,2017NewAR..78....1M,2018MNRAS.480.2011G,2018arXiv180909130M}.
If BH natal kicks are as high as the natal kicks inferred for neutron stars \citep{2001Natur.413..139M,2012MNRAS.425.2799R,2013MNRAS.434.1355J} from proper motion of pulsars \citep{2005MNRAS.360..974H}
then most BHs will escape the cluster. However, if BHs receive reduced kicks \citep{2002ApJ...572..407B,2005ApJ...625..324W,2006ApJ...650..303B,2012ApJ...747..111W} then depending on their final mass, which increases for lower metallicity progenitor stars, \citep{2002ApJ...567..532H,2009MNRAS.395L..71M,2010ApJ...714.1217B,2009MNRAS.400..677Z,2010MNRAS.408..234M,2012ApJ...749...91F,2013MNRAS.429.2298M,2015MNRAS.451.4086S,2017MNRAS.470.4739S} their BH retention fraction will be higher. Recently, \citet{2018MNRAS.481L.110B} has suggested that nucleosynthesis in accretion discs around retained stellar mass BHs in the early evolution of GCs could account for  present-day light element anticorrelations observed in GCs.

Even if BH retention fraction is high, the long-term survival of BHs in GCs has been debated in numerous studies. It had been postulated that BHs retained in GCs would segregate and form a dynamically decoupled subsystem where they would interact through strong encounters. This would result in their ejection from the GC within a billion years \citep{1993Natur.364..421K,1993Natur.364..423S,2000ApJ...528L..17P}. More recent numerical and theoretical works have shown that BH depletion might not be so efficient and in GCs with large initial half-mass relaxation times, a sizeable number of BHs can survive from few a Gyr up to a Hubble time and longer \citep{2013ApJ...763L..15M,2013MNRAS.430L..30S,2013MNRAS.432.2779B,2013MNRAS.436..584B,2014MNRAS.439.2459H,2015ApJ...800....9M,2015MNRAS.454.3150G,Wang2016,2016MNRAS.459.3432M,2016MNRAS.463.2109R,2018ApJ...855L..15K,2018MNRAS.479.4652A,2018MNRAS.478.1844A}. More works have also explored the impact of high BH retention on the dynamical evolution and observational properties of star clusters \citep{2004ApJ...608L..25M,2007MNRAS.379L..40M,2008MNRAS.386...65M,2010MNRAS.407.1946D,2010MNRAS.402..371B,2011ApJ...741L..12B,2014MNRAS.444...29L,2014MNRAS.441.3703Z,ngc6101,2016MNRAS.455...35A,2017IAUS..316..234C,2018MNRAS.474.3835W,2018ApJ...864...13W,2018A&A...617A..69P,2018MNRAS.481.5123B,2019MNRAS.482.4713Z,2019ApJ...871...38K}. 

Using global observational properties to determine whether a particular GC could contain a large number of BHs remains challenging. Results from GC simulations that retain a large number of BHs up to a Hubble time show that these clusters are typically characterized by relatively low central surface brightness values, large stellar core and half-light radii, and long half-mass relaxation times \citep{2018MNRAS.478.1844A}. However, such present-day observational properties are not unique to GC models with many BHs. Observational properties of GCs depend on their evolutionary history which is governed by their initial properties and GC models without too many BHs may also exhibit observational properties similar to models that sustain a large number of BHs. Building on the work presented in \citet{2018MNRAS.479.4652A} and \citet{2018MNRAS.478.1844A}, we experiment with several supervised machine learning classifiers that are trained on the observational properties of nearly 1300 simulated star cluster models surviving up to 12 Gyrs \citep{Askar2017}. The purpose of these classifiers is to use these observational properties to identify whether a GC could be harbouring a BH subsystem. We use these classifiers with readily available global observational properties of Galactic GCs (\citealt{harris1996}, \emph{updated 2010} and \citealt{holger}) to identify which of them could contain a BH system.

The rest of this paper proceeds as follows:
In \autoref{sec-gc-data}, we provide information on GC simulation models that were used in this study and which of their observational properties were selected in order to train the classifier. In \autoref{ml-c}, we provide details as to how the classifier was constructed and which machine learning algorithms were used. In \autoref{sec-predictions}, we apply the classifier we have developed to results from \textit{N}-body simulations and available observational data for Galactic GCs. The results are presented in \autoref{tab:harris} and \autoref{tab:holger}. In \autoref{high-conf}, we discuss results and pinpoint which Galactic GCs are most likely to contain many BHs and compare our results with previous studies. In Section \ref{sec:conclusions} we give the conclusions and provide links to access to our publicly available code for classifying GC models with a BH subsystem. 

\section{GC Simulation Models}\label{sec-gc-data}

For the purpose of this study, we used results from numerical simulations of GC models that were carried out using the
\textsc{mocca} code \citep{HG2013,Giersz2013} as part of the \textsc{MOCCA}-Survey Database I \citep{Askar2017} project. \textsc{mocca} is a code for simulating star clusters based on H\'{e}non's implementation of the Monte Carlo method \citep{Henon1971,std1982,Stod1986} to follow the long-term dynamical evolution of spherically symmetric stellar clusters. 

Additionally, the \textsc{mocca} code utilizes prescriptions for stellar and binary evolution from the SSE/BSE codes \citep{Hurley2000,Hurley2002} to evolve each star and binary system. For computing the outcome of strong interactions involving two binaries or a binary and single star, \textsc{mocca} uses the \textsc{fewbody} code \citep{Fregeau2004}. \textsc{MOCCA} also implements a realistic treatment for escapees in tidally limited clusters based on \citet{FH2000}. Results from the \textsc{mocca} code have been extensively compared with results from direct \textit{N}-body simulations of star clusters. There is a remarkable agreement in the evolution of global parameters and populations of specific objects \citep{Giersz2008,Giersz2013,Wang2016,Madrid2017}, however the \textsc{mocca} code is significantly faster than direct \textit{N}-body codes. The speed of the Monte Carlo method and the implementation of additional processes makes \textsc{mocca} an ideal tool to simulate the realistic evolution of a large set of GC models.\\

The \textsc{MOCCA}-Survey Database I \citep{Askar2017,Pasquato2018} project comprises nearly 2000 star cluster models that were simulated using the \textsc{mocca} code and span a wide range of initial conditions. The initial conditions of the simulated star cluster models are provided in Table 1 in \citep{Askar2017}. From these simulations about 1300 star cluster models survive up to 12 Gyrs (see \citealt{Pasquato2018} for initial properties of GC models that survive up to 12 Gyrs). In this paper, we make use of various 12 Gyr properties of these 1300 models as features. From these models, 162 models contain more than 15 BHs at 12 Gyrs. Following \citet{2018MNRAS.478.1844A}, these models are tagged as models with a BH subsystem (BHS). Models with fewer than 15 BHs are tagged as NO BHS models.

\subsection{Feature Selection and Analysis}

\revision{We use the following global properties to determine the presence of BH subsystems (in order of the amount of variance they explain in the PCA analysis):}

\begin{itemize}
    \item \revision{Central Surface Brightness}
    \item \revision{Central Velocity Dispersion}
    \item \revision{Total V-band Luminosity}
    \item \revision{Median Relaxation Time}
    \item \revision{Observational Half-Light Radius}
    \item \revision{Observational Core Radius}
\end{itemize}

In particular, these features were also chosen because these are common observed properties. For the Milky Way GCs, their values were readily available within two catalogues (\citealt{harris1996}, \emph{updated 2010} and \citealt{holger}). For the simulated GC models, snapshots from \textsc{mocca} simulations provide details of each star in the system including its position, radial and tangential velocities, luminosity, radii and magnitudes. Using this data, it is possible to compute the total V-band luminosity, surface brightness and velocity dispersion profiles for the cluster models. The Principal Componenent Analysis (PCA) as detailed in
\autoref{sec:pca} revealed these to be the features that explained
the most variance.

As detailed in \citet{2018MNRAS.478.1844A}, in order to calculate the central surface brightness from the simulated models we use the infinite projection method described in Appendix B of \citet{mashchenko2005} to generate a surface brightness profile for the GC at 12 Gyr; determine the central value in units of V-band luminosity per square pc. The profile is also used to obtain the observational core and half-light radii. \revision{The observational core radius is the projected radius at which the surface brightness is half its central value. The half-light radius is the projected radius which contains half of the total cluster light. To obtain the central velocity dispersion from the simulated GC models, we used the value in the innermost bin of the projected line-of-sight (LOS) velocity dispersion profile (made using only luminous stars brighter than $\rm M_{V}=6$ and with 50 logarithmic bins in radius) from the 12 Gyr projected snapshot.} The median relaxation time for the cluster models is computed using the total cluster luminosity and the half-light radius. The calculation is done in the same way as done for Galactic GCs in \citet[][\emph{updated 2010}]{harris1996}\footnote{See \url{http://physwww.mcmaster.ca/~harris/mwgc.ref} for details} using Equation 11 in \citep{1993ASPC...50..373D}:
\begin{equation}
t_{\rm rh} = 2.055\times10^{6} \rm yr \times \frac{1}{\ln(0.4N_{\star})}\langle m_{\star}\rangle^{-1} (M_{\rm cl})^{0.5} r_{\rm h}^{1.5}
\end{equation}
where $\rm M_{cl}$ ($\rm {M_{\odot}}$) is the cluster mass estimated using the cluster V-band luminosity and assuming a mean mass-to-light ratio of 2. $\langle \rm m_{\star} \rangle$ ($\rm {M_{\odot}}$) is the mean stellar mass which is assumed to be 1/3 $\rm {M_{\odot}}$ and $\rm N_{\star}$ is the total number of stars which is found by dividing $\rm M_{cl}$ by $\langle \rm m_{\star} \rangle$. Following \citet[][\emph{updated 2010}]{harris1996}, we assume that $\rm r_{h}$ is the half-light radius in units of pc. Therefore, the median relaxation time ($t_{\rm rh}$) essentially depends on the cluster luminosity and its half-light radius.
We also trained the classifier on using the proper half-mass relaxation time \citep[][Equation 2-62]{1987degc.book.....S} at 12 Gyr from the simulated cluster models instead of the median relaxation time. This was done so that we could use the classifier with the catalogue of Galactic GC parameters provided by \citet{holger} (see Section \ref{sec-predictions} for details).

\subsection{Principal Component Analysis}
\label{sec:pca}

\begin{figure}
	\centering
	\input{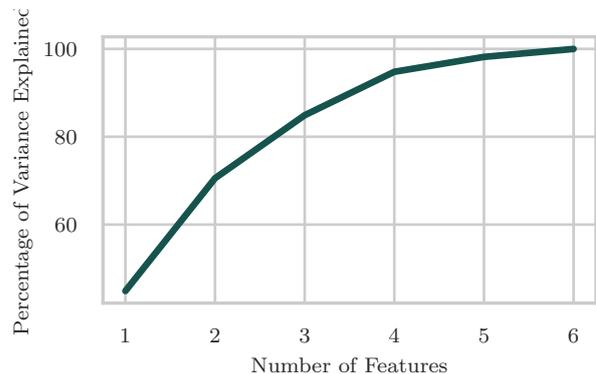}
    
	\caption{Number of features and the amount of variance they explain in the data              as given with PCA analysis.}
    \label{fig:PCA}
\end{figure}

We performed PCA as outlined by \cite{pca} on the data.
This allows us to gain insight into how many features are required to explain the
variance in the data in additional to figuring out the most important features.
The percentage of variance explained by the number of features is presented in
Figure \ref{fig:PCA}.

The following steps were applied before performing the PCA analysis. 
The $\log_{10}$ function was applied on the median relaxation time and central surface brightness.
All features were then normalized by centering to the mean and standardizing to a
unit variance.

\begin{figure*}
	\centering
    \includegraphics[scale=0.523]{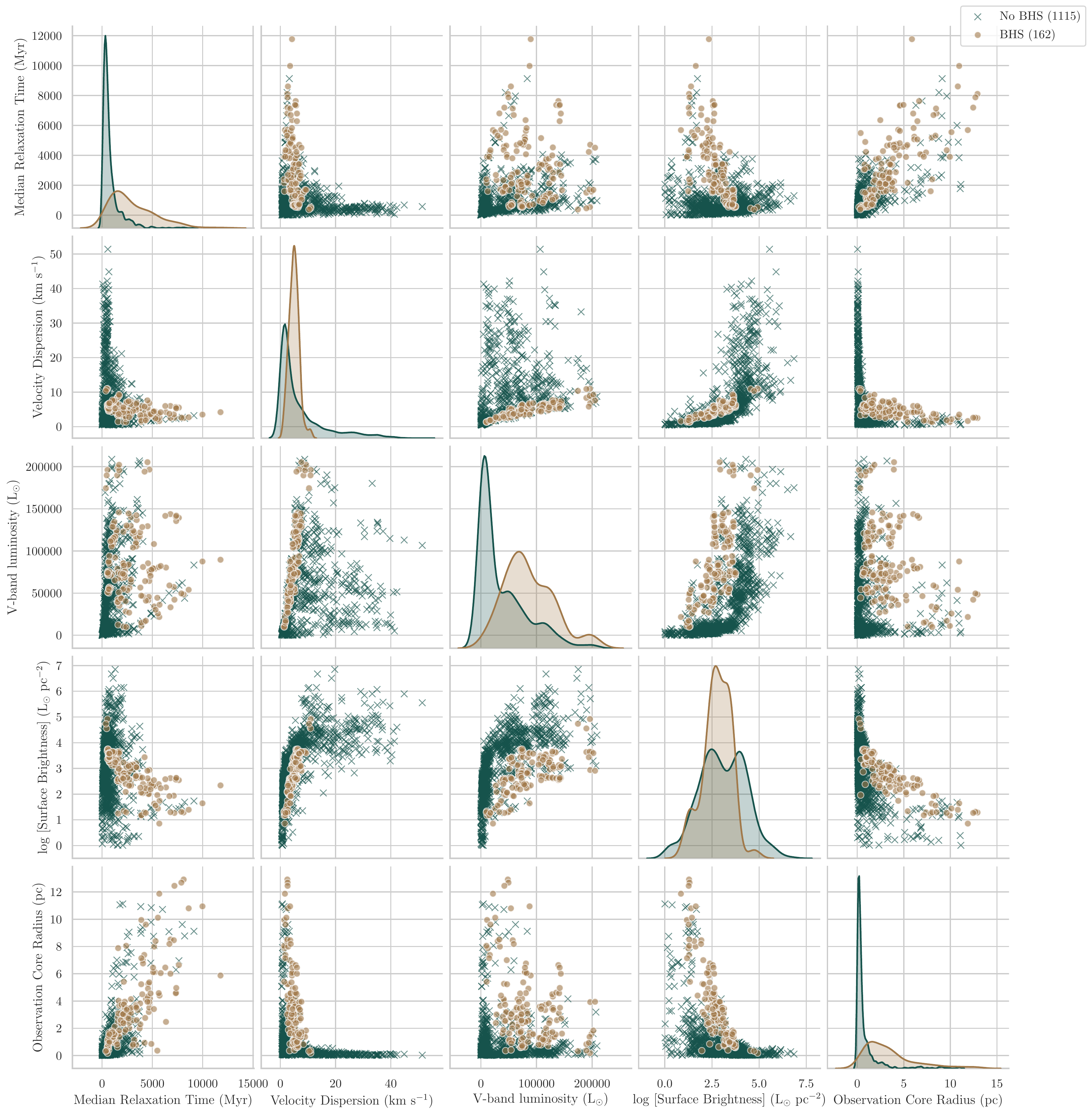}
    
    \caption{\revision{Pairwise plot of the 5 most important features. Each row and column
             represents a scatter plot between the two features. Matching rows and columns (diagonal panels) contain a distribution plot of the given observational parameter for cluster models with and without BH subsystems. A BHS containing cluster is defined as one that contains more than 15 BHs at 12 Gyr.}}
     \label{fig:pairplot}
\end{figure*}

Pairwise plots of these features are presented in Figure \ref{fig:pairplot}. There plots show some simple, visually discernible patterns that may be used to distinguish clusters with BH subsystems.

\section{Machine Learning Classifier}\label{ml-c}

The goal is to predict the presence of stellar mass BH subsystems (BHS) from the properties
of a GC. We use supervised learning algorithms, that is, a model is
trained upon data where we know whether a BHS is present or not. This model
then makes predictions on data where their presence is unknown. A learning process
tunes the model with the best possible parameters that minimize a cost function.

\subsection{\revision{Models Tested}}

We use \texttt{scikit-learn} \citep{scikit-learn} to evaluate different classification
methods as well as XGBoost \citep{xgboost} for the gradienst boosted tree implementation. An empirical analysis of each classifier is presented, in particular we were
after a model that offers good accuracy, minimizing false positives and one that we
could analyze to uncover insights into what makes BH subsystems more likely
to appear.

\subsubsection{Naive Bayes}

Naive Bayes classifiers are based on \citeauthor{Bayes}' theorem, using the 
training data to compute the conditional probabilities \citep{BayesClassifier}.
By computing the prior probabilities and assuming their independence, we can
compute the probability of a new data point $x$ and its likelihood to be a member
of a certain class $C$ with:

\begin{equation}
p(C \mid x) \propto p(C)\ p(x_1 \mid C) \ p(x_2 \mid C) \ \ldots
\end{equation}

where $x_1, x_2, \ldots$ are the features of the new data point $x$ and $C$ is
the predicted class. 

The naivity
in the name of the classifier is the assumption that the probabilities of each features
are independent, which is not necessarily true. In our case, one would expect, for
example, the
Central Surface Brightness and Core Radius of a GC to be correlated.
Hence, this classifier is mostly used as a baseline to evaluate the performance of
others.

\subsubsection{k-Nearest Neighbors}

$k$-Nearest Neighbors is a simple classification technique that relies on a
distance metric such as Euclidean distance or Manhattan distance and a
hyper-parameter $k$ explained below \citep{kNN}.

Intuitively, kNN computes the distance between a new data point $x$ with all
the points in the training set $x_i$. The $k$ points with the smallest distance
are considered to be ``closest'' in value and the average of their labels is
selected as the predicted label $y$ for $x$.

\subsubsection{Support Vector Machines}

\begin{figure}
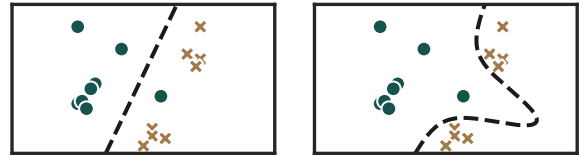

	\centering
	\input{plots/SVM_Linear.pgf}
    \input{plots/SVM_Poly.pgf}
    
    \caption{Example of linearity in SVM class boundaries with toy data. A less linear boundary may be able to fit the training data better but fail to generalize.}
    \label{fig:svm}
\end{figure}

Support vector machines attempt to compute a hyperplane that separates the 
input data points if they were to be plotted in $n$ dimensional space where 
$n$ is the number of features \citep{SVM}. 
We use hinge loss \citep{Hinge} which is defined as:

\begin{equation}
\max (0, 1 - y (\boldsymbol{w}^T \boldsymbol{x} + b))
\end{equation}

where $\boldsymbol{w}$ and $\boldsymbol{b}$ are the parameters of the hyperplane that need to be learned.
$y$ is the true class label, -1 if a BHS is not present or 1 if it is, $x$ are
the features for the particular example. These hyperplanes can vary in 
linearity as seen in Figure \ref{fig:svm}
potentially trading off over-fitting and generalization.

\subsubsection{Decision Trees}
\label{sec:decisiontrees}

Decision trees \citep{decisionTrees} are a powerful classification method that are
relatively simple but can yield very good results especially for simpler models. As
the name suggests, decision trees are a tree of ``decisions''.
These decisions could be, for example, whether the half light radius ($R_e$) is more
than 2pc.
The learning process finds decisions that split the input data in an attempt to
minimize a loss function. 

In particular, we use the Gini impurity \citep{GINI} as the cost function:

\begin{equation}
1 - \sum_{j=1}^{c}{p^2_j}
\end{equation}

where $p_j$ is the probability of a particular class being chosen at a given split.
Thus, a split that completely divides the input set into two particular classes
would result in the probability of one class being $1$ and the other being $0$,
resulting in the most minimum score of 0.

These splits are computed at every level of the tree until a max depth is reached,
a certain Gini score is achieved or some other heuristic. The scores are computed
at each decision making this a greedy algorithm.

One major advantage of decision trees is that they are not a complete black box,
they can easily be introspected to reveal why the model makes certain decisions.
This can be a very powerful tool to analyze the insights that the model might
have generated.

\subsubsection{Gradient Boosted Decision Trees}

Gradient boosted decision trees attempt to improve on two of the major shortcomings
of regular decision trees. Firstly, their tendency to overfit and secondly,
offering a departure from a completely greedy approach.

Specifically, gradient boosting involves training dozens to hundreds of 
shallow decision trees and then aggregating their outputs with a 
set of weights to create a final prediction \citep{GradientBoost}. Each tree is
made to focus on the ``mistakes'' of previous trees by weighing samples that were
mislabelled. Thus, the final classifier looks along the lines of:

\begin{equation}
f(x) = \sum_{i = 1}^{N}{\gamma_i\ g_i(x)}
\end{equation}

where $N$ is the number of shallow trees trained, $g_i$ is the output of the $i$th
shallow tree and $\gamma_i$ is the weight computed for it that minimizes the loss
function.

Gradient boosted trees offer better accuracy than plain old decision trees because
they can generate far more complex models \citep{GradientBoost2}. On the other 
hand, this makes them much more of a black box. It is hard to investigate and 
determine exactly why a certain prediction is being made one way or the other.

\subsection{Scoring Metric And Testing}

Classifiers were evaluated using stratified k-fold testing. Briefly, k-fold cross
validation involves splitting up the training data into $k$ overlapping subsets,
within each of these subsets a portion of the data is used for testing and another
portion for validation.

Stratified k-fold testing involves generating subsets that have a similar
representation of classes as the input data set. If the original data has 30\%
true labels, then each subset will be chosen to have around 30\% true values as
well.
As noted by \cite{kohavi1995study}, this can result in far better bias and 
variance numbers when performing cross-validation.

Within each cross-validation test, a particular metric must be chosen to evaluate
the performance of each classifier. In particular, we want a binary
classification metric, since the problem we were trying to solve is whether
a BH subsystem is present or not. A metric that is insensitive to imbalanced
data is essential, since only around 12\% of the simulated clusters have BH
subsystems.

We also preferred metrics that weigh false-positives more aggressively than false-negatives. 
This allows us to single out good candidate clusters that can then be studied observationally to
verify the presence of BH subsystems. Thus, we want to be very sure that
the clusters classified are very likely to have a BH subsystem.

The metrics we looked at were:
\begin{itemize}
	\item \textbf{Precision} \citep{Recall}
    		\begin{equation}
            \text{Precision} = \frac{tp}{tp + fp}
            \end{equation}
            where $tp$ is the number of true positives and $fp$ is the number of false positives.
            Precision is the fraction of positive classifications that
            were actually correct. This metric ensures that we will be very
            confident if a BH subsystem is predicted, but we won't
            necessarily be able to find all of them.
            \vspace{1mm}
            
	\item \textbf{ROC-AUC} \citep{ROC} 

			\vspace{1mm}
    		The Receiving Operating Characteristic is a plot of the ratio of
            true positives and false positives. The area under this curve is an
            indicator of the classifier's performance. The ROC curve is affected
            by true-negatives, which is problematic with our imbalanced data 
            \citep{davis2006relationship}.
            \vspace{1mm}
            
	\item \textbf{F-Score} \citep{FScore}
 
    		\vspace{1mm}
    		The F-Score is the harmonic mean of precision and recall. Recall 
            \citep{han2011data} is defined as
            \begin{equation}
            \text{Recall} = \frac{tp}{tp + fn}
            \end{equation}
            where $tp$ is the number of true positives and $fn$ is the number of false negatives.
            Recall ensures that we correctly classify all positive examples. 
            Thus with f-score, we are confident there are not many false positives
            and that we manage to find all the clusters with BH subsystems.
\end{itemize}

\begin{figure}
	\centering
    \input{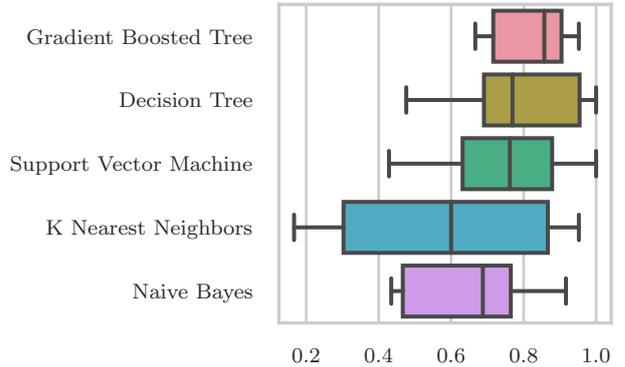}
    
    \caption{Comparison of each classifiers' f-score with 15-fold testing. An f-score of 1.0 is the best possible, 0.0 is the worst. The score is affected by how well the classifier can find all the BH subsystems and whether the identifications are false-positives.}
    \label{fig:comparison}
\end{figure}

F-Score was chosen as the metric to use in our situation since it aligns best with
the objectives we noted above. In Figure \ref{fig:comparison} we present a comparison
of the f-score of each classifier.

From the results, it is apparent that the decision trees offer the best performance
characteristics. The gradient boosted trees have a slightly higher f-score and a
smaller spread than regular decision trees. Hence, moving forward we decided to use
both kinds of decision trees, offering results from the better performing gradient
boosted trees and introspecting and presenting insights from the regular decision
trees. \revision{In particular, decision trees were chosen because they are
relatively easy to inspect and analyze why a given classification decision
was made. This does however mean that the analysis in \autoref{sec:analysisTree}
is sensitive to the order of branches taken. That is, the features must be
decided upon in the sequence presented in the tree or else predictions
will not be accurate. Gradient boosted trees were
used for their high accuracy as they conglomerate several shallow trees:
every individual split in the smaller trees has smaller effect on the final classification \citep{freund1999short}.}

\subsection{Tuning Hyperparameters and Analysing Classifiers}

Machine learning algorithms rely on certain parameters which are not learned,
but are instead provided by the user. For decision trees, this includes properties 
such as the max depth of the tree, the minimum change in the cost function and
the number of trees used in gradient boosting. These hyperparameters are often
tuned by hand based on heuristics about the data or they can be optimized 
computationally.

We used a technique known as ``Random Searching'' by \cite{bergstra2012random}
in order to find the optimal hyperparameters. We used this over an exhaustive
search such as a grid search \citep{bergstra2011algorithms} because it is much
less computationally expensive.

Random searching involves randomly sampling values for each hyperparameter from
a range, such as a max depth of a decision tree between 5 and 20. The classifier
is then trained using these hyperparameters and $k$-fold testing is performed to
evaluate which set of hyperparameters provide the best performance according to
our scoring metric, f-score.

\begin{figure}
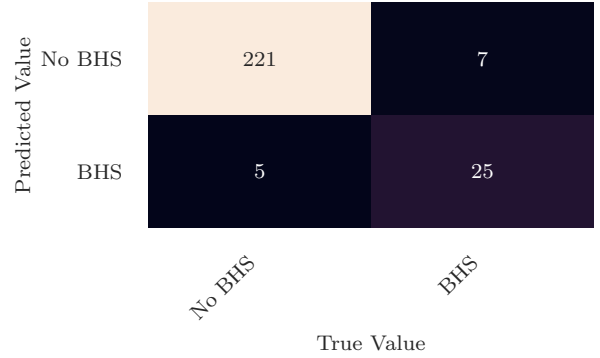

	\centering
\begingroup%
\makeatletter%
\begin{pgfpicture}%
\pgfpathrectangle{\pgfpointorigin}{\pgfqpoint{3.102947in}{1.917727in}}%
\pgfusepath{use as bounding box, clip}%
\begin{pgfscope}%
\pgfsetbuttcap%
\pgfsetmiterjoin%
\definecolor{currentfill}{rgb}{1.000000,1.000000,1.000000}%
\pgfsetfillcolor{currentfill}%
\pgfsetlinewidth{0.000000pt}%
\definecolor{currentstroke}{rgb}{1.000000,1.000000,1.000000}%
\pgfsetstrokecolor{currentstroke}%
\pgfsetdash{}{0pt}%
\pgfpathmoveto{\pgfqpoint{0.000000in}{0.000000in}}%
\pgfpathlineto{\pgfqpoint{3.102947in}{0.000000in}}%
\pgfpathlineto{\pgfqpoint{3.102947in}{1.917727in}}%
\pgfpathlineto{\pgfqpoint{0.000000in}{1.917727in}}%
\pgfpathclose%
\pgfusepath{fill}%
\end{pgfscope}%
\begin{pgfscope}%
\pgfsetbuttcap%
\pgfsetmiterjoin%
\definecolor{currentfill}{rgb}{1.000000,1.000000,1.000000}%
\pgfsetfillcolor{currentfill}%
\pgfsetlinewidth{0.000000pt}%
\definecolor{currentstroke}{rgb}{0.000000,0.000000,0.000000}%
\pgfsetstrokecolor{currentstroke}%
\pgfsetstrokeopacity{0.000000}%
\pgfsetdash{}{0pt}%
\pgfpathmoveto{\pgfqpoint{0.710500in}{0.655983in}}%
\pgfpathlineto{\pgfqpoint{3.019614in}{0.655983in}}%
\pgfpathlineto{\pgfqpoint{3.019614in}{1.834394in}}%
\pgfpathlineto{\pgfqpoint{0.710500in}{1.834394in}}%
\pgfpathclose%
\pgfusepath{fill}%
\end{pgfscope}%
\begin{pgfscope}%
\definecolor{textcolor}{rgb}{0.150000,0.150000,0.150000}%
\pgfsetstrokecolor{textcolor}%
\pgfsetfillcolor{textcolor}%
\pgftext[x=0.972487in,y=0.169464in,left,base,rotate=45.000000]{\color{textcolor}\rmfamily\fontsize{8.000000}{9.600000}\selectfont No BHS}%
\end{pgfscope}%
\begin{pgfscope}%
\definecolor{textcolor}{rgb}{0.150000,0.150000,0.150000}%
\pgfsetstrokecolor{textcolor}%
\pgfsetfillcolor{textcolor}%
\pgftext[x=2.259116in,y=0.301536in,left,base,rotate=45.000000]{\color{textcolor}\rmfamily\fontsize{8.000000}{9.600000}\selectfont BHS}%
\end{pgfscope}%
\begin{pgfscope}%
\definecolor{textcolor}{rgb}{0.150000,0.150000,0.150000}%
\pgfsetstrokecolor{textcolor}%
\pgfsetfillcolor{textcolor}%
\pgftext[x=1.865057in,y=0.098667in,,top]{\color{textcolor}\rmfamily\fontsize{8.000000}{9.600000}\selectfont True Value}%
\end{pgfscope}%
\begin{pgfscope}%
\definecolor{textcolor}{rgb}{0.150000,0.150000,0.150000}%
\pgfsetstrokecolor{textcolor}%
\pgfsetfillcolor{textcolor}%
\pgftext[x=0.154222in,y=1.501235in,left,base]{\color{textcolor}\rmfamily\fontsize{8.000000}{9.600000}\selectfont No BHS}%
\end{pgfscope}%
\begin{pgfscope}%
\definecolor{textcolor}{rgb}{0.150000,0.150000,0.150000}%
\pgfsetstrokecolor{textcolor}%
\pgfsetfillcolor{textcolor}%
\pgftext[x=0.341000in,y=0.912030in,left,base]{\color{textcolor}\rmfamily\fontsize{8.000000}{9.600000}\selectfont BHS}%
\end{pgfscope}%
\begin{pgfscope}%
\definecolor{textcolor}{rgb}{0.150000,0.150000,0.150000}%
\pgfsetstrokecolor{textcolor}%
\pgfsetfillcolor{textcolor}%
\pgftext[x=0.098667in,y=1.245188in,,bottom,rotate=90.000000]{\color{textcolor}\rmfamily\fontsize{8.000000}{9.600000}\selectfont Predicted Value}%
\end{pgfscope}%
\begin{pgfscope}%
\pgfpathrectangle{\pgfqpoint{0.710500in}{0.655983in}}{\pgfqpoint{2.309114in}{1.178410in}}%
\pgfusepath{clip}%
\pgfsetbuttcap%
\pgfsetroundjoin%
\definecolor{currentfill}{rgb}{0.981377,0.920617,0.865369}%
\pgfsetfillcolor{currentfill}%
\pgfsetlinewidth{0.000000pt}%
\definecolor{currentstroke}{rgb}{1.000000,1.000000,1.000000}%
\pgfsetstrokecolor{currentstroke}%
\pgfsetdash{}{0pt}%
\pgfpathmoveto{\pgfqpoint{0.710500in}{1.834394in}}%
\pgfpathlineto{\pgfqpoint{1.865057in}{1.834394in}}%
\pgfpathlineto{\pgfqpoint{1.865057in}{1.245188in}}%
\pgfpathlineto{\pgfqpoint{0.710500in}{1.245188in}}%
\pgfpathlineto{\pgfqpoint{0.710500in}{1.834394in}}%
\pgfusepath{fill}%
\end{pgfscope}%
\begin{pgfscope}%
\pgfpathrectangle{\pgfqpoint{0.710500in}{0.655983in}}{\pgfqpoint{2.309114in}{1.178410in}}%
\pgfusepath{clip}%
\pgfsetbuttcap%
\pgfsetroundjoin%
\definecolor{currentfill}{rgb}{0.018319,0.022977,0.107385}%
\pgfsetfillcolor{currentfill}%
\pgfsetlinewidth{0.000000pt}%
\definecolor{currentstroke}{rgb}{1.000000,1.000000,1.000000}%
\pgfsetstrokecolor{currentstroke}%
\pgfsetdash{}{0pt}%
\pgfpathmoveto{\pgfqpoint{1.865057in}{1.834394in}}%
\pgfpathlineto{\pgfqpoint{3.019614in}{1.834394in}}%
\pgfpathlineto{\pgfqpoint{3.019614in}{1.245188in}}%
\pgfpathlineto{\pgfqpoint{1.865057in}{1.245188in}}%
\pgfpathlineto{\pgfqpoint{1.865057in}{1.834394in}}%
\pgfusepath{fill}%
\end{pgfscope}%
\begin{pgfscope}%
\pgfpathrectangle{\pgfqpoint{0.710500in}{0.655983in}}{\pgfqpoint{2.309114in}{1.178410in}}%
\pgfusepath{clip}%
\pgfsetbuttcap%
\pgfsetroundjoin%
\definecolor{currentfill}{rgb}{0.010608,0.018082,0.100187}%
\pgfsetfillcolor{currentfill}%
\pgfsetlinewidth{0.000000pt}%
\definecolor{currentstroke}{rgb}{1.000000,1.000000,1.000000}%
\pgfsetstrokecolor{currentstroke}%
\pgfsetdash{}{0pt}%
\pgfpathmoveto{\pgfqpoint{0.710500in}{1.245188in}}%
\pgfpathlineto{\pgfqpoint{1.865057in}{1.245188in}}%
\pgfpathlineto{\pgfqpoint{1.865057in}{0.655983in}}%
\pgfpathlineto{\pgfqpoint{0.710500in}{0.655983in}}%
\pgfpathlineto{\pgfqpoint{0.710500in}{1.245188in}}%
\pgfusepath{fill}%
\end{pgfscope}%
\begin{pgfscope}%
\pgfpathrectangle{\pgfqpoint{0.710500in}{0.655983in}}{\pgfqpoint{2.309114in}{1.178410in}}%
\pgfusepath{clip}%
\pgfsetbuttcap%
\pgfsetroundjoin%
\definecolor{currentfill}{rgb}{0.135016,0.075856,0.190441}%
\pgfsetfillcolor{currentfill}%
\pgfsetlinewidth{0.000000pt}%
\definecolor{currentstroke}{rgb}{1.000000,1.000000,1.000000}%
\pgfsetstrokecolor{currentstroke}%
\pgfsetdash{}{0pt}%
\pgfpathmoveto{\pgfqpoint{1.865057in}{1.245188in}}%
\pgfpathlineto{\pgfqpoint{3.019614in}{1.245188in}}%
\pgfpathlineto{\pgfqpoint{3.019614in}{0.655983in}}%
\pgfpathlineto{\pgfqpoint{1.865057in}{0.655983in}}%
\pgfpathlineto{\pgfqpoint{1.865057in}{1.245188in}}%
\pgfusepath{fill}%
\end{pgfscope}%
\begin{pgfscope}%
\definecolor{textcolor}{rgb}{0.150000,0.150000,0.150000}%
\pgfsetstrokecolor{textcolor}%
\pgfsetfillcolor{textcolor}%
\pgftext[x=1.287778in,y=1.539791in,,]{\color{textcolor}\rmfamily\fontsize{8.000000}{9.600000}\selectfont 221}%
\end{pgfscope}%
\begin{pgfscope}%
\definecolor{textcolor}{rgb}{1.000000,1.000000,1.000000}%
\pgfsetstrokecolor{textcolor}%
\pgfsetfillcolor{textcolor}%
\pgftext[x=2.442335in,y=1.539791in,,]{\color{textcolor}\rmfamily\fontsize{8.000000}{9.600000}\selectfont 7}%
\end{pgfscope}%
\begin{pgfscope}%
\definecolor{textcolor}{rgb}{1.000000,1.000000,1.000000}%
\pgfsetstrokecolor{textcolor}%
\pgfsetfillcolor{textcolor}%
\pgftext[x=1.287778in,y=0.950586in,,]{\color{textcolor}\rmfamily\fontsize{8.000000}{9.600000}\selectfont 5}%
\end{pgfscope}%
\begin{pgfscope}%
\definecolor{textcolor}{rgb}{1.000000,1.000000,1.000000}%
\pgfsetstrokecolor{textcolor}%
\pgfsetfillcolor{textcolor}%
\pgftext[x=2.442335in,y=0.950586in,,]{\color{textcolor}\rmfamily\fontsize{8.000000}{9.600000}\selectfont 25}%
\end{pgfscope}%
\end{pgfpicture}%
\makeatother%
\endgroup%
    
    \caption{\revision{Confusion matrix for decision tree classifier. The classifier was trained on a random 75\% portion of the data that was sampled with stratification. Predicted values were then tested for the remaining 25\% against the real values, each value's categorization is then presented here.}}
    \label{fig:confusiondecision}
\end{figure}

\begin{figure}
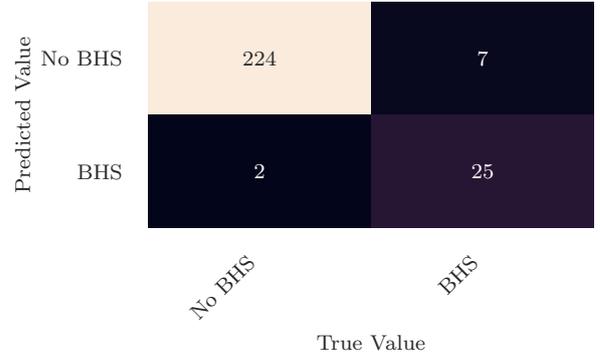

	\centering
\begingroup%
\makeatletter%
\begin{pgfpicture}%
\pgfpathrectangle{\pgfpointorigin}{\pgfqpoint{3.102947in}{1.917727in}}%
\pgfusepath{use as bounding box, clip}%
\begin{pgfscope}%
\pgfsetbuttcap%
\pgfsetmiterjoin%
\definecolor{currentfill}{rgb}{1.000000,1.000000,1.000000}%
\pgfsetfillcolor{currentfill}%
\pgfsetlinewidth{0.000000pt}%
\definecolor{currentstroke}{rgb}{1.000000,1.000000,1.000000}%
\pgfsetstrokecolor{currentstroke}%
\pgfsetdash{}{0pt}%
\pgfpathmoveto{\pgfqpoint{0.000000in}{0.000000in}}%
\pgfpathlineto{\pgfqpoint{3.102947in}{0.000000in}}%
\pgfpathlineto{\pgfqpoint{3.102947in}{1.917727in}}%
\pgfpathlineto{\pgfqpoint{0.000000in}{1.917727in}}%
\pgfpathclose%
\pgfusepath{fill}%
\end{pgfscope}%
\begin{pgfscope}%
\pgfsetbuttcap%
\pgfsetmiterjoin%
\definecolor{currentfill}{rgb}{1.000000,1.000000,1.000000}%
\pgfsetfillcolor{currentfill}%
\pgfsetlinewidth{0.000000pt}%
\definecolor{currentstroke}{rgb}{0.000000,0.000000,0.000000}%
\pgfsetstrokecolor{currentstroke}%
\pgfsetstrokeopacity{0.000000}%
\pgfsetdash{}{0pt}%
\pgfpathmoveto{\pgfqpoint{0.710500in}{0.655983in}}%
\pgfpathlineto{\pgfqpoint{3.019614in}{0.655983in}}%
\pgfpathlineto{\pgfqpoint{3.019614in}{1.834394in}}%
\pgfpathlineto{\pgfqpoint{0.710500in}{1.834394in}}%
\pgfpathclose%
\pgfusepath{fill}%
\end{pgfscope}%
\begin{pgfscope}%
\definecolor{textcolor}{rgb}{0.150000,0.150000,0.150000}%
\pgfsetstrokecolor{textcolor}%
\pgfsetfillcolor{textcolor}%
\pgftext[x=0.972487in,y=0.169464in,left,base,rotate=45.000000]{\color{textcolor}\rmfamily\fontsize{8.000000}{9.600000}\selectfont No BHS}%
\end{pgfscope}%
\begin{pgfscope}%
\definecolor{textcolor}{rgb}{0.150000,0.150000,0.150000}%
\pgfsetstrokecolor{textcolor}%
\pgfsetfillcolor{textcolor}%
\pgftext[x=2.259116in,y=0.301536in,left,base,rotate=45.000000]{\color{textcolor}\rmfamily\fontsize{8.000000}{9.600000}\selectfont BHS}%
\end{pgfscope}%
\begin{pgfscope}%
\definecolor{textcolor}{rgb}{0.150000,0.150000,0.150000}%
\pgfsetstrokecolor{textcolor}%
\pgfsetfillcolor{textcolor}%
\pgftext[x=1.865057in,y=0.098667in,,top]{\color{textcolor}\rmfamily\fontsize{8.000000}{9.600000}\selectfont True Value}%
\end{pgfscope}%
\begin{pgfscope}%
\definecolor{textcolor}{rgb}{0.150000,0.150000,0.150000}%
\pgfsetstrokecolor{textcolor}%
\pgfsetfillcolor{textcolor}%
\pgftext[x=0.154222in,y=1.501235in,left,base]{\color{textcolor}\rmfamily\fontsize{8.000000}{9.600000}\selectfont No BHS}%
\end{pgfscope}%
\begin{pgfscope}%
\definecolor{textcolor}{rgb}{0.150000,0.150000,0.150000}%
\pgfsetstrokecolor{textcolor}%
\pgfsetfillcolor{textcolor}%
\pgftext[x=0.341000in,y=0.912030in,left,base]{\color{textcolor}\rmfamily\fontsize{8.000000}{9.600000}\selectfont BHS}%
\end{pgfscope}%
\begin{pgfscope}%
\definecolor{textcolor}{rgb}{0.150000,0.150000,0.150000}%
\pgfsetstrokecolor{textcolor}%
\pgfsetfillcolor{textcolor}%
\pgftext[x=0.098667in,y=1.245188in,,bottom,rotate=90.000000]{\color{textcolor}\rmfamily\fontsize{8.000000}{9.600000}\selectfont Predicted Value}%
\end{pgfscope}%
\begin{pgfscope}%
\pgfpathrectangle{\pgfqpoint{0.710500in}{0.655983in}}{\pgfqpoint{2.309114in}{1.178410in}}%
\pgfusepath{clip}%
\pgfsetbuttcap%
\pgfsetroundjoin%
\definecolor{currentfill}{rgb}{0.981377,0.920617,0.865369}%
\pgfsetfillcolor{currentfill}%
\pgfsetlinewidth{0.000000pt}%
\definecolor{currentstroke}{rgb}{1.000000,1.000000,1.000000}%
\pgfsetstrokecolor{currentstroke}%
\pgfsetdash{}{0pt}%
\pgfpathmoveto{\pgfqpoint{0.710500in}{1.834394in}}%
\pgfpathlineto{\pgfqpoint{1.865057in}{1.834394in}}%
\pgfpathlineto{\pgfqpoint{1.865057in}{1.245188in}}%
\pgfpathlineto{\pgfqpoint{0.710500in}{1.245188in}}%
\pgfpathlineto{\pgfqpoint{0.710500in}{1.834394in}}%
\pgfusepath{fill}%
\end{pgfscope}%
\begin{pgfscope}%
\pgfpathrectangle{\pgfqpoint{0.710500in}{0.655983in}}{\pgfqpoint{2.309114in}{1.178410in}}%
\pgfusepath{clip}%
\pgfsetbuttcap%
\pgfsetroundjoin%
\definecolor{currentfill}{rgb}{0.032852,0.030888,0.118630}%
\pgfsetfillcolor{currentfill}%
\pgfsetlinewidth{0.000000pt}%
\definecolor{currentstroke}{rgb}{1.000000,1.000000,1.000000}%
\pgfsetstrokecolor{currentstroke}%
\pgfsetdash{}{0pt}%
\pgfpathmoveto{\pgfqpoint{1.865057in}{1.834394in}}%
\pgfpathlineto{\pgfqpoint{3.019614in}{1.834394in}}%
\pgfpathlineto{\pgfqpoint{3.019614in}{1.245188in}}%
\pgfpathlineto{\pgfqpoint{1.865057in}{1.245188in}}%
\pgfpathlineto{\pgfqpoint{1.865057in}{1.834394in}}%
\pgfusepath{fill}%
\end{pgfscope}%
\begin{pgfscope}%
\pgfpathrectangle{\pgfqpoint{0.710500in}{0.655983in}}{\pgfqpoint{2.309114in}{1.178410in}}%
\pgfusepath{clip}%
\pgfsetbuttcap%
\pgfsetroundjoin%
\definecolor{currentfill}{rgb}{0.010608,0.018082,0.100187}%
\pgfsetfillcolor{currentfill}%
\pgfsetlinewidth{0.000000pt}%
\definecolor{currentstroke}{rgb}{1.000000,1.000000,1.000000}%
\pgfsetstrokecolor{currentstroke}%
\pgfsetdash{}{0pt}%
\pgfpathmoveto{\pgfqpoint{0.710500in}{1.245188in}}%
\pgfpathlineto{\pgfqpoint{1.865057in}{1.245188in}}%
\pgfpathlineto{\pgfqpoint{1.865057in}{0.655983in}}%
\pgfpathlineto{\pgfqpoint{0.710500in}{0.655983in}}%
\pgfpathlineto{\pgfqpoint{0.710500in}{1.245188in}}%
\pgfusepath{fill}%
\end{pgfscope}%
\begin{pgfscope}%
\pgfpathrectangle{\pgfqpoint{0.710500in}{0.655983in}}{\pgfqpoint{2.309114in}{1.178410in}}%
\pgfusepath{clip}%
\pgfsetbuttcap%
\pgfsetroundjoin%
\definecolor{currentfill}{rgb}{0.152013,0.081591,0.202705}%
\pgfsetfillcolor{currentfill}%
\pgfsetlinewidth{0.000000pt}%
\definecolor{currentstroke}{rgb}{1.000000,1.000000,1.000000}%
\pgfsetstrokecolor{currentstroke}%
\pgfsetdash{}{0pt}%
\pgfpathmoveto{\pgfqpoint{1.865057in}{1.245188in}}%
\pgfpathlineto{\pgfqpoint{3.019614in}{1.245188in}}%
\pgfpathlineto{\pgfqpoint{3.019614in}{0.655983in}}%
\pgfpathlineto{\pgfqpoint{1.865057in}{0.655983in}}%
\pgfpathlineto{\pgfqpoint{1.865057in}{1.245188in}}%
\pgfusepath{fill}%
\end{pgfscope}%
\begin{pgfscope}%
\definecolor{textcolor}{rgb}{0.150000,0.150000,0.150000}%
\pgfsetstrokecolor{textcolor}%
\pgfsetfillcolor{textcolor}%
\pgftext[x=1.287778in,y=1.539791in,,]{\color{textcolor}\rmfamily\fontsize{8.000000}{9.600000}\selectfont 224}%
\end{pgfscope}%
\begin{pgfscope}%
\definecolor{textcolor}{rgb}{1.000000,1.000000,1.000000}%
\pgfsetstrokecolor{textcolor}%
\pgfsetfillcolor{textcolor}%
\pgftext[x=2.442335in,y=1.539791in,,]{\color{textcolor}\rmfamily\fontsize{8.000000}{9.600000}\selectfont 7}%
\end{pgfscope}%
\begin{pgfscope}%
\definecolor{textcolor}{rgb}{1.000000,1.000000,1.000000}%
\pgfsetstrokecolor{textcolor}%
\pgfsetfillcolor{textcolor}%
\pgftext[x=1.287778in,y=0.950586in,,]{\color{textcolor}\rmfamily\fontsize{8.000000}{9.600000}\selectfont 2}%
\end{pgfscope}%
\begin{pgfscope}%
\definecolor{textcolor}{rgb}{1.000000,1.000000,1.000000}%
\pgfsetstrokecolor{textcolor}%
\pgfsetfillcolor{textcolor}%
\pgftext[x=2.442335in,y=0.950586in,,]{\color{textcolor}\rmfamily\fontsize{8.000000}{9.600000}\selectfont 25}%
\end{pgfscope}%
\end{pgfpicture}%
\makeatother%
\endgroup%
    
    \caption{Confusion matrix for gradient boosted tree classifier. The generation methodology is the same as for Figure \ref{fig:confusiondecision}.}
    \label{fig:confusiongradient}
\end{figure}

The search results for ideal hyperparameters \revision{for the Harris data
set} are presented in Table \ref{tab:hyperdecision} and Table \ref{tab:hypergradient},
for decision trees and gradient boosted decision trees respectively.
\revision{Hyperparameters for the \citet{holger} dataset were searched seperately
due to differences in relaxation time calculation.}
Confusion
matrices \citep{confusionmatrix} are presented in Figure \ref{fig:confusiondecision} and Figure
\ref{fig:confusiongradient}. The confusion matrices offer a promising result,
the amount of false positives is low for both the plain old decision trees and the
gradient boosted trees.

\begin{table}
	\caption{Ideal hyperparameters found for the decision tree classifier using
             random search with 15-fold verification with an f1 score of 
             $0.798 \pm 0.11$ }
    
    \centering
	\begin{tabular}{lp{3.6cm}r}
    	\toprule
    	Hyperparameter & Description & Value \\
        \midrule
        Max Depth & Maximum number of nodes from the root node down to a leaf of the tree. & 5 \\
        Loss Function & Function to maximize for each split, either entropy gain or gini impurity decrease. & Gini \\
        Max Features & The random subset of features to use when calculating splits. Used to reduce overfitting. & 5 \\
        Min Split Sample & The minimum number of samples during a decision before creating a split. & 2 \\
        Min Impurity Decrease & The minimum decrease in the loss function to create a split in the tree. & 0.05 \\
        \bottomrule
    \end{tabular}
             
	\label{tab:hyperdecision}
\end{table}

\begin{table}
	\caption{Ideal hyperparameters found for gradient boosted decision tree
             classifier using random search with 15-fold verification with an f1
             score of $0.857 \pm 0.14$}
             
    \centering
	\begin{tabular}{lp{3.6cm}r}
    	\toprule
    	Hyperparameter & Description & Value \\
        \midrule
        Loss Function & Function to minimize when choosing gradients and tree weights. & Mean Precision \\
        Subsample & Percentage of samples to train base decision trees on. & 100\% \\
        Estimators & The number of base decision trees to ensemble together. & 850 \\
        Max Depth & Maximum number of nodes from the root down a to a leaf in the decision trees. & 6 \\
        Learning Rate & The factor to multiply the gradients by when performing gradient descent search. & 0.3 \\
        Base Score & Initial weight for each training instance, the global bias. & 0.5 \\
        \bottomrule
    \end{tabular}    
             
    \label{tab:hypergradient}
\end{table}

\subsection{Analysis of Decision Tree}
\label{sec:analysisTree}

A full visualization of the decision tree is presented in Figure \ref{fig:decisiontree}.
The first branching of the decision tree is based on core radius.
Clusters with a small observational core radius are very unlikely to host BH
subsystems, and this is reflected by the fact that $29$ out of $963$
($3\%$) of the clusters with $r_c < 1.15$ pc host a BH subsystem, as
opposed to 162 out of 1289 for the whole simulation sample ($12\%$).
From a physical point of view, this is due to the presence of a BH subsystem which provides energy to surrounding luminous stars and prevents them from segregating in order to form a bright core.
This results in a larger core radius value and lower central surface brightness. Following the bottom branch of the tree, where BH
subsystem hosts are more abundant ($133$ out of $326$, i.e.  $37\%$),
the next split is on total V-band luminosity. Simulated clusters with
total luminosity above $2.6 \times 10^4 L_\odot$, which is a typical
value for a small cluster such as NGC 2298, are more than ten times as
likely to be BH subsystem hosts than smaller clusters (host fraction
$69\%$ compared to $6\%$). This result is also straightforward
to interpret, as high present-day luminosity implies larger initial mass which increases the number of BHs that form in the cluster. A larger number of BHs can be retained in these clusters which also typically have longer initial relaxation times. 
Even if initial half-mass relaxation time is short for massive and dense clusters, they may still have a sizeable population of BHs at 12 Gyr. This is because BH retention fraction in the first 30 Myrs is higher due to large escape velocity from the dense cluster. Therefore, despite being dynamically older, such clusters may still contain comparable number of BHs at 12 Gyr to a dynamically younger GC model with a longer initial half-mass relaxation time. Further branches show that high
central velocity dispersion is also an indicator of higher probability
of being a BH subsystem host and so is low central surface brightness
and long relaxation time, which are also expected based on our
physical understanding of the dynamical effects of a BH subsystem \citep{2018MNRAS.479.4652A}.

\section{Predictions Using Machine Learning}\label{sec-predictions}

\subsection{Predictions on results from \textit{N}-body simulations}

Before applying the classifier on available data for present-day structural parameters and properties available for Galactic GCs, we tested its efficacy by checking if it could determine the presence of a BH subsystem in GC models simulated with a direct \textit{N}-body code instead of \textsc{mocca}. \citet{Wang2016} used the \textsc{nbody6++gpu} \citep{wang2} to simulate the evolution of four GC models with a million initial stars. The three models D1-R7-IMF93, D2-R7-IMF01 and D3-R7-ROT in \citet{Wang2016} were simulated up to 12 Gyr, at that time they had 245, 1037 and 1096 BHs respectively. We took the observational properties with which the classifier was trained from the 12 Gyr data available for those three models. The properties and the results from the classifier are shown in Table \ref{tab:dragoon}. For all the three GC models, the classifiers correctly predicted the presence of a BHS. This includes both the classifier that was trained on all \textsc{mocca}-Survey I  GC models that survive up to 12 Gyr and the one that was trained only on simulated results in which mass fallback was enabled and BH kicks were lower.

\begin{table*}
	\caption{Predictions for the presence of BH subsystems with 12 Gyr data taken from the DRAGON million star GC direct \textit{N}-body simulations \citep{Wang2016} carried out using \textsc{nbody6++gpu} \citep{wang2}}
    
    \centering
    
    \begin{tabular}{lllllllll}
    \toprule
    Simulation & Half-Light & Central Surface & CVD \textsuperscript{$a$} & Total V-band & MRT \textsuperscript{$b$} & OCR \textsuperscript{$c$} & BHS & BHS \\
    Name & Radius (pc) & Brightness ($\text{L}_{\odot}$ $\text{pc}^{-2}$) & (km $\text{s}^{-1}$) & Luminosity ($\text{L}_{\odot}$) & (Myr) & (pc) & Prediction & (Fallback) \\
    \midrule
	D1-R7-IMF93 & 8.7 & $2.5 \times 10^2$ & 4.5 & $1.86 \times 10^5$ & 7417 & 4.8 & \textbf{True} & \textbf{True} \\
    D2-R7-IMF01 & 14.4 & $7.0 \times 10^1$ & 3.8 & $1.11 \times 10^5$ & 12706 & 10.0 & \textbf{True} & \textbf{True} \\
    D3-R7-ROT & 13.4 & $1.3 \times 10^2$ & 4.0 & $1.22 \times 10^5$ & 11874 & 11.9 & \textbf{True} & \textbf{True} \\
    \bottomrule
    \end{tabular}
    
    {\footnotesize
    \textsuperscript{$a$}Central Velocity Dispersion \quad\textsuperscript{$b$}Median Relaxation Time \quad\textsuperscript{$c$}Observational Core Radius
    }
             
	\label{tab:dragoon}
\end{table*}

\subsection{Predictions on observational data for Galactic GCs}

Having tested the trained classifier model described in the previous section on GC models simulated by \citet{Wang2016} using a direct \textit{N}-body code, we then predicted whether
stellar mass BHs were present in Galactic GCs using global observable properties from the
\citet[][\emph{updated 2010}]{harris1996} catalogue. This includes total V-band luminosity, half-light and core radius, central surface brightness, median relaxation time and central velocity dispersion. Similar to \citet{2018MNRAS.478.1844A}, we applied the classifier to Galactic GCs that have Galactocentric radii smaller than 17 kpc. This was done because nearly 99 per cent of the GC models that had a BHS subsystem at 12 Gyr
had Galactocentric radii smaller than 17 kpc. Distant Galactic GCs are not taken into account, this is because of the limited number of GC models with large Galactocentric distances in the \textsc{mocca}-Survey Database I models. Only three initial tidal radii values were taken in \textsc{mocca}-Survey Database I \citep{Askar2017} and due to these limited initial parameters, no GC models with initial number of objects larger than 100,000 had Galactocentric radii larger than 20 kpc. The \citet[][\emph{updated 2010}]{harris1996} catalogue does not provide central velocity dispersion for all Galactic GCs. For Galactic GCs in \citet[][\emph{updated 2010}]{harris1996} where central velocity dispersion data was not available, we re-trained and used the classifier without this feature. One classifier was trained on all \textsc{mocca}-Survey I models that survived up to 12 Gyr and another classifier was trained on models in which mass fallback was enabled and BH kicks were lower. The observational properties that were used and the results from the classifiers for \citet[][\emph{updated 2010}]{harris1996} catalogue data are provided in Table \ref{tab:harris}.

Additionally, we also used data for Galactic GC parameters provided by \citet{holger}\footnote{Data for GC parameters from \citet{holger} is available online at: \url{https://people.smp.uq.edu.au/HolgerBaumgardt/globular/} [Retrieved August 23, 2018]}. These parameters were obtained for 112 Galactic GCs by fitting a large grid of direct \textit{N}-body simulations to the observed velocity dispersion and surface density profiles. In order to use our classifier on the data provided by \citet{holger}, we converted cluster mass and central surface mass density \footnote{See \url{https://people.smp.uq.edu.au/HolgerBaumgardt/globular/combined_table.txt} [Retrieved August 23, 2018]} to total V-band luminosity and central surface brightness using the mass-to-light ratio that they provided. The central surface brightness estimated from this data is an upper value as mass-to-light ratio is expected to be higher in the central part of the cluster due to segregated massive compact remnants. Through their fitting procedure, \citet{holger} were also able to determine the half-mass relaxation time for the Galactic GCs in their catalogue. The classifier which was applied to data from \citet{holger} catalogue was thus trained on the half-mass relaxation time at 12 Gyr in our simulated cluster models instead of the median relaxation time (the classifier used for \citet[][\emph{updated 2010}]{harris1996} catalogue data was trained on the median relaxation time). GC parameters and the results from the classifier for the \citet{holger} data are provided in Table \ref{tab:holger}.

Galactic GCs that were classified as having a BH subsystem in either one of the two catalogues (using either classifier trained on all models or only on fallback models) have been listed in Table \ref{tab:checkmarks}. The green check marks indicate that the GC was classified as having a BHS. 
The columns indicate classification results from the two catalogues (\citet[][\emph{updated 2010}]{harris1996} (marked as Harris) or \citet{holger} (marked as B\&H)). Additionally, we distinguish where the classifier was trained on all models (BHS) and where it was trained only on models where mass fallback was enabled (Fallback). 
We have also indicated in \autoref{tab:checkmarks} which of the clusters had been identified by \cite{2018MNRAS.478.1844A} as having a BH subsystem. Clusters which did not have a central velocity dispersion value in the \citet[][\emph{updated 2010}]{harris1996} catalogue have also been indicated. 

\newcommand{\smalldag}{\!\!\! *}
\newcommand{\smallmark}{$\dagger$}

\begin{table}
	\centering
    \caption{\revision{Predictions from the \citet[][\emph{updated 2010}]{harris1996} and \citet{holger} datasets using the gradient boosted decision tree classifier. Entires where BHS presence was classified positively are shown. The \emph{BHS} column represents the classifier trained on all simulation data whereas \emph{Fallback} represents training on models where mass fallback was enabled and BH natal kicks were lower.}}
    \begin{tabular}{|l|cccc|}
    \toprule
    Cluster Name & 
    \scriptsize BHS & 
    \scriptsize Fallback & 
    \scriptsize BHS & 
    \scriptsize Fallback \\
                 & \scriptsize (Harris)       
                 & \scriptsize (Harris)       
                 & \scriptsize (B\&H) 
                 & \scriptsize (B\&H) \\
    \midrule
IC 4499 \smalldag & \cellcolor{green!25} \cmark & \cellcolor{green!25} \cmark & \cellcolor{green!25} \cmark & \cellcolor{green!25} \cmark \\
NGC 288 \smalldag & \cellcolor{green!25} \cmark & \cellcolor{green!25} \cmark & \cellcolor{green!25} \cmark & \cellcolor{green!25} \cmark \\
NGC 3201 \smalldag & \cellcolor{green!25} \cmark & \cellcolor{green!25} \cmark & \cellcolor{green!25} \cmark & \cellcolor{green!25} \cmark \\
NGC 4372 \smalldag \smallmark & \cellcolor{green!25} \cmark & \cellcolor{green!25} \cmark & \cellcolor{green!25} \cmark & \cellcolor{green!25} \cmark \\
NGC 4590 (M68) & \cellcolor{red!25} \xmark & \cellcolor{green!25} \cmark & \cellcolor{red!25} \xmark & \cellcolor{red!25} \xmark \\
NGC 4833 \smalldag \smallmark & \cellcolor{red!25} \xmark & \cellcolor{green!25} \cmark & \cellcolor{green!25} \cmark & \cellcolor{green!25} \cmark \\
NGC 5139 ($\omega$ Cen) & \cellcolor{green!25} \cmark & \cellcolor{green!25} \cmark & \cellcolor{red!25} \xmark & \cellcolor{green!25} \cmark \\
NGC 5272 (M3) \smalldag & \cellcolor{red!25} \xmark & \cellcolor{green!25} \cmark & \cellcolor{red!25} \xmark & \cellcolor{green!25} \cmark \\
NGC 5286 & \cellcolor{red!25} \xmark & \cellcolor{green!25} \cmark & \cellcolor{red!25} \xmark & \cellcolor{green!25} \cmark \\
NGC 5466 \smalldag & \cellcolor{red!25} \xmark & \cellcolor{green!25} \cmark & \cellcolor{red!25} \xmark & \cellcolor{red!25} \xmark \\
NGC 5897 \smalldag \smallmark & \cellcolor{green!25} \cmark & \cellcolor{green!25} \cmark & \cellcolor{red!25} \xmark & \cellcolor{green!25} \cmark \\
NGC 5904 (M5) & \cellcolor{red!25} \xmark & \cellcolor{green!25} \cmark & \cellcolor{red!25} \xmark & \cellcolor{green!25} \cmark \\
NGC 5927 & \cellcolor{red!25} \xmark & \cellcolor{red!25} \xmark & \cellcolor{green!25} \cmark & \cellcolor{green!25} \cmark \\
NGC 5986 \smalldag \smallmark & \cellcolor{green!25} \cmark & \cellcolor{green!25} \cmark & \cellcolor{red!25} \xmark & \cellcolor{green!25} \cmark \\
NGC 6101 \smalldag \smallmark & \cellcolor{green!25} \cmark & \cellcolor{green!25} \cmark & \cellcolor{red!25} \xmark & \cellcolor{red!25} \xmark \\
NGC 6139 \smallmark & \cellcolor{green!25} \cmark & \cellcolor{green!25} \cmark & \cellcolor{red!25} \xmark & \cellcolor{red!25} \xmark \\
NGC 6144 \smalldag \smallmark & \cellcolor{red!25} \xmark & \cellcolor{green!25} \cmark & \cellcolor{red!25} \xmark & \cellcolor{green!25} \cmark \\
NGC 6205 (M13) \smalldag & \cellcolor{red!25} \xmark & \cellcolor{green!25} \cmark & \cellcolor{green!25} \cmark & \cellcolor{green!25} \cmark \\
NGC 6218 (M12) & \cellcolor{green!25} \cmark & \cellcolor{green!25} \cmark & \cellcolor{red!25} \xmark & \cellcolor{red!25} \xmark \\
NGC 6254 (M10) & \cellcolor{green!25} \cmark & \cellcolor{green!25} \cmark & \cellcolor{red!25} \xmark & \cellcolor{green!25} \cmark \\
NGC 6266 (M62) & \cellcolor{red!25} \xmark & \cellcolor{red!25} \xmark & \cellcolor{green!25} \cmark & \cellcolor{red!25} \xmark \\
NGC 6273 (M19) \smallmark & \cellcolor{red!25} \xmark & \cellcolor{green!25} \cmark & \cellcolor{red!25} \xmark & \cellcolor{green!25} \cmark \\
NGC 6287 \smallmark & \cellcolor{green!25} \cmark & \cellcolor{green!25} \cmark & \cellcolor{red!25} \xmark & \cellcolor{red!25} \xmark \\
NGC 6304 \smallmark & \cellcolor{red!25} \xmark & \cellcolor{green!25} \cmark & \cellcolor{red!25} \xmark & \cellcolor{green!25} \cmark \\
NGC 6316 \smallmark & \cellcolor{green!25} \cmark & \cellcolor{green!25} \cmark & \cellcolor{red!25} \xmark & \cellcolor{red!25} \xmark \\
NGC 6333 (M9) \smallmark & \cellcolor{green!25} \cmark & \cellcolor{green!25} \cmark & \cellcolor{red!25} \xmark & \cellcolor{red!25} \xmark \\
NGC 6356 \smallmark & \cellcolor{red!25} \xmark & \cellcolor{green!25} \cmark & \cellcolor{red!25} \xmark & \cellcolor{green!25} \cmark \\
NGC 6362 \smalldag & \cellcolor{red!25} \xmark & \cellcolor{green!25} \cmark & \cellcolor{red!25} \xmark & \cellcolor{green!25} \cmark \\
NGC 6380 \smallmark & \cellcolor{green!25} \cmark & \cellcolor{green!25} \cmark & \cellcolor{red!25} \xmark & \cellcolor{red!25} \xmark \\
NGC 6388 & \cellcolor{red!25} \xmark & \cellcolor{red!25} \xmark & \cellcolor{red!25} \xmark & \cellcolor{green!25} \cmark \\
NGC 6401 \smalldag & \cellcolor{green!25} \cmark & \cellcolor{green!25} \cmark & \cellcolor{red!25} \xmark & \cellcolor{red!25} \xmark \\
NGC 6402 (M14) \smallmark & \cellcolor{green!25} \cmark & \cellcolor{green!25} \cmark & \cellcolor{green!25} \cmark & \cellcolor{green!25} \cmark \\
NGC 6426 \smalldag \smallmark & \cellcolor{red!25} \xmark & \cellcolor{green!25} \cmark & \cellcolor{red!25} \xmark & \cellcolor{red!25} \xmark \\
NGC 6440 \smallmark & \cellcolor{green!25} \cmark & \cellcolor{green!25} \cmark & \cellcolor{red!25} \xmark & \cellcolor{red!25} \xmark \\
NGC 6496 \smalldag \smallmark & \cellcolor{red!25} \xmark & \cellcolor{green!25} \cmark & \cellcolor{red!25} \xmark & \cellcolor{red!25} \xmark \\
NGC 6517 \smallmark & \cellcolor{red!25} \xmark & \cellcolor{green!25} \cmark & \cellcolor{red!25} \xmark & \cellcolor{red!25} \xmark \\
NGC 6539 (GCL 85) & \cellcolor{green!25} \cmark & \cellcolor{green!25} \cmark & \cellcolor{red!25} \xmark & \cellcolor{red!25} \xmark \\
NGC 6553 & \cellcolor{green!25} \cmark & \cellcolor{green!25} \cmark & \cellcolor{red!25} \xmark & \cellcolor{red!25} \xmark \\
NGC 6569 \smalldag \smallmark & \cellcolor{green!25} \cmark & \cellcolor{green!25} \cmark & \cellcolor{green!25} \cmark & \cellcolor{green!25} \cmark \\
NGC 6584 \smalldag \smallmark & \cellcolor{green!25} \cmark & \cellcolor{green!25} \cmark & \cellcolor{red!25} \xmark & \cellcolor{red!25} \xmark \\
NGC 6656 (M22) \smalldag & \cellcolor{green!25} \cmark & \cellcolor{green!25} \cmark & \cellcolor{red!25} \xmark & \cellcolor{green!25} \cmark \\
NGC 6712 \smalldag & \cellcolor{red!25} \xmark & \cellcolor{green!25} \cmark & \cellcolor{green!25} \cmark & \cellcolor{green!25} \cmark \\
NGC 6723 \smalldag \smallmark & \cellcolor{green!25} \cmark & \cellcolor{green!25} \cmark & \cellcolor{green!25} \cmark & \cellcolor{green!25} \cmark \\
NGC 6760 \smallmark & \cellcolor{green!25} \cmark & \cellcolor{green!25} \cmark & \cellcolor{red!25} \xmark & \cellcolor{red!25} \xmark \\
NGC 6779 (M56) \smalldag & \cellcolor{red!25} \xmark & \cellcolor{green!25} \cmark & \cellcolor{green!25} \cmark & \cellcolor{green!25} \cmark \\
NGC 6809 (M55) \smalldag & \cellcolor{red!25} \xmark & \cellcolor{green!25} \cmark & \cellcolor{green!25} \cmark & \cellcolor{green!25} \cmark \\
NGC 6934 \smalldag & \cellcolor{green!25} \cmark & \cellcolor{green!25} \cmark & \cellcolor{red!25} \xmark & \cellcolor{red!25} \xmark \\
NGC 6981 (M72) \smalldag & \cellcolor{red!25} \xmark & \cellcolor{green!25} \cmark & \cellcolor{red!25} \xmark & \cellcolor{red!25} \xmark \\
NGC 7078 (M15) & \cellcolor{red!25} \xmark & \cellcolor{red!25} \xmark & \cellcolor{green!25} \cmark & \cellcolor{green!25} \cmark \\
NGC 7089 (M2) & \cellcolor{red!25} \xmark & \cellcolor{green!25} \cmark & \cellcolor{red!25} \xmark & \cellcolor{red!25} \xmark \\
Pal11 & \cellcolor{green!25} \cmark & \cellcolor{green!25} \cmark & \cellcolor{green!25} \cmark & \cellcolor{green!25} \cmark \\
Terzan5 \smallmark & \cellcolor{red!25} \xmark & \cellcolor{green!25} \cmark & \cellcolor{red!25} \xmark & \cellcolor{red!25} \xmark \\
	\bottomrule
    \end{tabular}
    \vspace{1mm}
    {\footnotesize
    
    * Clusters identified to contain BHS by \citet{2018MNRAS.478.1844A}
    
    $\dagger\ $ Clusters without Central Velocity Dispersion data in Harris catalogue.
    }
    \label{tab:checkmarks}
\end{table}

\section{Results - Milky Way GCs Likely to Contain
BH Subsystem}\label{high-conf}

Using the gradient boosted decision tree classifier and data from two different catalogues, we were able to shortlist 52 Galactic GCs that may contain a BH subsystem in Table \ref{tab:checkmarks}. Out of these 52 clusters, 18 were identified in both \citet[][\emph{updated 2010}]{harris1996} and \citet{holger} data sets. Among the 52 GCs that were identified by the machine learning classifier as having a BH subsystem, 27 of them had been previously identified by \citet{2018MNRAS.478.1844A} as having a BH subsystem. In their study, \citet{2018MNRAS.478.1844A} had identified 29 Galactic GCs that could contain a BH subsystem by manually comparing observational properties of \textsc{mocca}-Survey I models that had many BHs at 12 Gyr with the observed properties of Galactic GCs. Our automated approach is able to recover their findings and further identify additional GCs that could possibly contain a sizeable number of BHs. There were only 2 Galactic GCs identified by \citet{2018MNRAS.478.1844A} as having a BH subsystem that were not classified as having a BH subsystem in this study. \revision{These were NGC 6171 (M 107) and IC 1276 (Pal 7). Both NGC 6171 and IC 1276 have low central surface brightness and sufficiently long median relaxation time, however their V-band luminosity, observational core and half-light radii are not as large as other GCs identified as having a BH subsystem. While the relaxation time and surface brightness were determined to be the most important features in the PCA analysis from \autoref{sec:pca}, the
V-band luminosity and radii values are sufficient to sway the classifier to
a negative classification.}

\revision{8 GCs were classified as having a BH subsystem with the classifier trained on all models as well as fallback models for both \citet[][\emph{updated 2010}]{harris1996} and \citet{holger} catalogues. These clusters were IC 4499, NGC 288, NGC 3201, NGC 4372, NGC 6402 (M14), NGC 6569, NGC 6723 and Pal 11. Among these GCs, a BH candidate has been observed through radial velocity measurements in NGC 3201 \citep{2018MNRAS.475L..15G}}.

After NGC 5139 ($\rm \omega$ Cen), IC 4499 has the second largest half-mass relaxation time among all Galactic GCs within 17 kpc from the Galactic center in the \citet{holger} catalogue. The cluster has an unusually large core and half-light radii values that are consistent with GC models that contain a large number of BHs at a Hubble time. Among all Galactic GCs, IC 4499 central surface brightness to total V-band luminosity ratio is one of the smallest. \citet{2018MNRAS.478.1844A} estimated that IC 4499 could contain few hundreds of BHs. 

NGC 288 is another cluster which could contain a large population of BHs. \citet{2016MNRAS.462.1937S} used extensive spectroscopic and photometric survey data to estimate the fraction of non-luminous mass within the half-light radius of NGC 288 and NGC 6218 (M12). They found that about 60 per cent of the mass inside the half-light radius could be in dark remnants. It may be possible that a significant amount of this unseen mass is in stellar mass BHs.

NGC 4372 is an old metal poor GC in the Galactic halo which has large core and half-light radii compared to most Galactic GCs. \citep{2014A&A...567A..69K} carried out a detailed kinematic observations of NGC 4372 and found evidence for systematic rotation. They also obtained a half-light radius value lower than that reported in \citet[][\emph{updated 2010}]{harris1996} and also provided the central velocity dispersion value. Using the parameters that they derived, the classifier still identified NGC 4372 is having a BH subsystem. Interestingly, \citet{2008A&A...480..397S} found 9 accreting X-ray sources in NGC 4372 and none of them were inside the half-light radius. Such close binary systems are expected to more centrally segregated due to their larger mass. Lack of segregation of such systems could point towards a population of more massive dark remnants in the central parts of NGC 4372. 

Other Galactic GCs which were classified as having a BH subsystem in at least 3 columns shown in Table \ref{tab:checkmarks} were NGC 4833, NGC 5139 ($\omega$ Cen), NGC 5897, NGC 6205 (M13), NGC 6254 (M10), NGC 6656 (M22), NGC 6712 and NGC 6779 (M56). Among these clusters, accreting BH candidates have been observed in NGC 6656 (M22) and NGC 6254 (M10) \citep{2012Natur.490...71S,2018ApJ...855...55S}. Other accreting BH candidates have been observed in NGC 6266 (M62) and NGC 104 (47 Tuc). M62 has been classified as having a BH subsystem in Table \ref{tab:checkmarks} for parameters provided in the \citet{holger} catalogue. However, it \revision{is} not classified as having a BH subsystem when we use the parameters provided in the \citet[][\emph{updated 2010}]{harris1996} catalogue. M62's total V-band luminosity in the \citet{holger} catalogue is almost half the value given in \citet[][\emph{updated 2010}]{harris1996} catalogue. Also its half-mass relaxation time ($\sim 1350$ Myr) in \citet{holger} is longer than the median relaxation time ($\sim 955$ Myr) given in the \citet[][\emph{updated 2010}]{harris1996} catalogue. Both M10 and M62 have small core and half-light radii compared to most GC models that have a BH subsystem. This could indicate that these GCs are depleting their BH population and evolving towards a second core bounce \citep{2013MNRAS.435.1047B}. During this phase of the cluster evolution, the size of the BH subsystem decreases while its density increases. This can lead to stronger dynamical interactions and it has been shown that a higher fraction of BHs in such clusters are in binaries and some of these could be mass transferring systems \citep{2018ApJ...852...29K,2018MNRAS.479.4652A,2018MNRAS.478.1844A}. Interestingly, M10 was considered as a possible candidate intermediate-mass black hole host by \citet{2010ApJ...713..194B}, even though their results were inconclusive.

\begin{figure}
	\centering
	\input{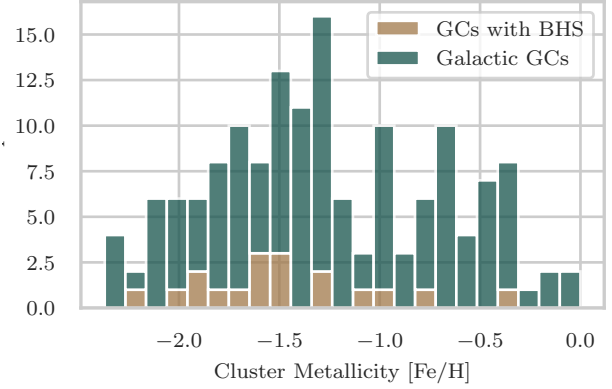}
    
    \caption{\revision{Distribution of cluster metallicities for all Milky Way GCs taken from the \citet[][\emph{updated 2010}]{harris1996} catalogue in green. Metallicity distribution for Galactic GCs classified as having a BH subsystem with at least 3 ticks in \autoref{tab:checkmarks} are are labeled with a different color.}}
    \label{fig:metal}
\end{figure}

\begin{table}
  \centering
  \caption{\revision{Galactic GCs classified as having a BH subsystem with at least 3 ticks in \autoref{tab:checkmarks} along with their metallicities.}}
	\begin{tabular}{lc}
    	\toprule
        Cluster Name & Metallicity [Fe/H] \\
        \midrule
        IC 4499 \smalldag & -1.53 \\
        NGC 288 \smalldag & -1.32 \\
        NGC 3201 \smalldag & -1.59 \\
        NGC 4372 \smalldag & -2.17 \\
        NGC 4833 \smalldag & -1.85 \\
        NGC 5139 ($\omega$ Cen) & -1.53 \\
        NGC 5897 \smalldag & -1.9 \\
        NGC 5986 \smalldag & -1.59 \\
        NGC 6205 (M13) \smalldag & -1.53 \\
        NGC 6254 (M10) & -1.56 \\
        NGC 6402 (M14) & -1.28 \\
        NGC 6569 \smalldag & -0.76 \\
        NGC 6656 (M22) \smalldag & -1.7 \\
        NGC 6712 \smalldag & -1.02 \\
        NGC 6723 \smalldag & -1.1 \\
        NGC 6779 (M56) \smalldag & -1.98 \\
        NGC 6809 (M55) \smalldag & -1.94 \\
        Pal 11 \smalldag & -0.4 \\
        \bottomrule
    \end{tabular}
    
    \vspace{1mm}
    {\footnotesize
    
    * Clusters identified to contain BHS by \citet{2018MNRAS.478.1844A}
    }
    \label{tab:finallist}
\end{table}

In Figure \ref{fig:metal}, we compared the metallicity distribution for the GCs that were classified as having a BH Subsystem for parameters provided in both \citet[][\emph{updated 2010}]{harris1996} and \cite{holger} catalogues. \revision{Most of these GCs are on the low metallicity end for GCs in the Milky Way, however, it cannot be conclusively stated from Figure \ref{fig:metal} that BHS are exclusively present in low metallicty GCs}. Metallicity plays an important role in the evolution of massive stars that are progenitors of BHs. Lower metallicity stars lose less mass due to stellar winds and produce more massive BHs \citep{2001A&A...369..574V,2016A&A...594A..97B, 2017MNRAS.470.4739S}. If BH natal kicks depend on BH masses, then BHs that have low metallicity progenitors will have lower natal kicks. This can increase their retention fraction in GCs and therefore present-day GCs that are most likely to contain many BHs should have low metallicities. Moreover, if low metallicity GCs form more massive BHs then the BH subsystem can keep the cluster in a state of balanced evolution \citep{2013MNRAS.436..584B} for a longer time as massive BHs in binaries burn out slower \citep{2018MNRAS.479.4652A}. This can slow down the depletion of the BHs and delay the second core bounce \citep{2013MNRAS.436..584B}. In Table \ref{tab:finallist}, we provide a list of 18 Galactic GCs that are most likely to contain a BH subsystem based on ML classifier results on their available parameters in both \citet[][\emph{updated 2010}]{harris1996} and \cite{holger} catalogues. The metallicities of these GCs taken from \citet[][\emph{updated 2010}]{harris1996} are also shown in Table \ref{tab:finallist}.

\subsection{\revision{Limitations and Caveats}}

It is important to point out that our results and the training of the classifiers significantly depends on how realistic our GC models are and how accurate are the observed parameters for Galactic GCs. While the simulated GC models take into account the most important physical processes that drive the dynamical evolution of GCs, there are other factors that could change present-day observational properties of GC models. This includes cluster rotation, proper treatment of the external tidal field and taking into account the clusters orbit (see \citealt{2018MNRAS.478.1844A} for a detailed discussion on limitations of using GC simulations to infer information about observed GCs). 

For such a classifier to work properly, one needs strong constraints on the observed global properties of GCs. While many nearby Galactic GCs have been well studied, there can be many disagreements between values for distance, total luminosity, half-light and core radii between different studies. Obtaining parameters like central velocity dispersion or surface brightness through observations can be very challenging and usually these observed values can have significant errors. \revision{Moreover, it is important to point out that observational parameters obtained from simulated models may not take into account the limitations and difficulties that plague observations of Galactic GCs. For instance, the velocity dispersion profile obtained from our simulated models is constructed taking into account the line-of-sight velocity of stars brighter than 2 magnitudes below the turn-off magnitude, where as velocity dispersion profiles obtained from spectroscopic observations of GCs are constructed using bright subgiant and giant stars. This can lead to difficulties in obtaining the actual central velocity dispersion of a GC. However, this issue is shared by any procedure that relies on comparing simulated and observed quantities, so it is not a weakness of our machine learning approach \emph{per se}.}

Results from the classifier can be sensitive to the values provided for the observable properties that were used as training features. For example, in Table \ref{tab:checkmarks}, NGC 7078 (M15) is classified as having a BH subsystem for parameters provided for it in the \citet{holger} catalogue. This is a dense core collapsed cluster with a high central surface brightness value ($\rm 1 \times 10^{5} \rm L_{\odot} pc^{-2}$) in the \citet[][\emph{updated 2010}]{harris1996} catalogue and one would not expect the presence of a BH subsystem in such a cluster. However, in the
\citet{holger} catalogue, its central mass surface density ($\rm log(\Sigma_{c})$
in units of $\rm M_{\odot} pc^{-2}$) was just 0.45\footnote{\url{https://people.smp.uq.edu.au/HolgerBaumgardt/globular/combined_table.txt}} which resulted in it being classified as having a BH subsystem. Therefore, it is important to have correct values for such parameters in order for the classifier to properly work. Similarly, Terzan 5 is classified as having a BH subsystem for \citet[][\emph{updated 2010}]{harris1996} data with the classifier trained only on models in which mass fallback was enabled. The GC has a compact dense core and a short median relaxation time. Therefore, it is not likely to contain a large number of BHs. However, its central surface brightness in \citet[][\emph{updated 2010}]{harris1996} catalogue is about 3 orders of magnitude lower than the one in the \citet{holger} catalogue and for this reason it is classified as having a BH subsystem. For clusters like Terzan 5 which have a large colour excess ($\rm E(B-V)\lesssim 2.82$ \citep{2012ApJ...755L..32M}), estimates of central surface brightness can have large errors. Moreover, the central surface brightness values provided in the \citet[][\emph{updated 2010}]{harris1996} catalogue are mostly from ground-based imaging data, with spatial resolution of about 1 arcsecond\footnote{\url{http://physwww.mcmaster.ca/~harris/mwgc.ref}} and there could be large uncertainty in these values. With future telescopes and data releases from the \textit{Gaia} mission, better observational data will be available for Galactic GCs. This will provide strong constraints on GC parameters which will be useful in determining whether they could contain a large population of BHs.  

Apart from uncertainties in observational data, it is also important to point out the limitations in the simulated GC models. For \textsc{mocca}-Survey Database I GC models, the evolution of massive stars and final BH masses were based on prescriptions provided in the \textsc{sse/bse} \citep{Hurley2000,Hurley2002} codes. BH natal kicks were drawn from a Maxwellian distribution (with $\rm \sigma=265 km s^{-1}$) \citep{2005MNRAS.360..974H} and in the fallback models, the kick magnitude depends on the BH mass (these were obtained using the fallback prescription for remnant masses from \citet{2002ApJ...572..407B}). Recent developments in wind prescriptions \citep{2011A&A...535A..56G}, supernovae models \citep{2012ApJ...749...91F}, evolution of massive stars (including pair and pulsational instability supernovae \citep{2016A&A...594A..97B,2017MNRAS.470.4739S}) \revision{and binary systems \citep{2018arXiv180904605S,2019arXiv190100863D} can strongly influence final BH masses and retention fractions in GCs.} The mass function of retained BHs in GCs can subsequently affect the evolution of the BH subsystem and how it shapes the global properties of the cluster. Therefore, results from future simulations with improved physics for evolution of BH progenitors will be useful in better determining present-day properties of GCs that retain a large number of BHs. Moreover, other important factors such as distribution of the initial binary parameters \citep{2017MNRAS.471.2812B}, common envelope evolution \citep{2017MNRAS.468.2429B,2018MNRAS.480.2011G} can also be important in determining compact object populations and present-day observational properties of GC models. \revision{As discussed above, there might be significant differences in the number of retained BHs and their masses between the current training set (simulated GC models) and the prediction set (GCs in our Galaxy). However, the purpose of this study is to simply classify whether an observed Galactic GC contains a BH subsystem using a classifier that gives the lowest f-score. Therefore, these differences should not strongly influence the results and may only be relevant for GCs that are border-line cases.}  







































\section{Conclusions}\label{sec:conclusions}

In this work, we applied machine learning in a new and novel way to 
astrophysics: predicting real world properties based on simulations.
An empirical comparison was
performed between multiple classifiers with the guiding metric
primarily being a low false positive rate (f-score). We successfully trained a
fairly accurate gradient boosted decision tree classifier on simulation data and then applied
the learned model onto real world data. $K$-fold testing with our
simulation data showed that the model had an accuracy of 95\% and a
false positive rate of less than 1\%. Moreover, applying the classifier
on an independent set of simulations ($N$-body vs Monte Carlo) yielded
accurate results as well, a classifier trained on \textsc{mocca} Monte Carlo simulations
was able to accurately predict the presence of BHS in $N$-body \citep{Wang2016} simulated
clusters using just the observational properties.

Applied onto real world data, our results are also fairly
encouraging. We ran our model on two different catalogues of 
Milky Way GCs in order to get predictions based on independent data sets. We managed to successfully identify several clusters which 
have been ascertained to contain BHs by previous observational
studies. This includes the likes of NGC 3201 as corroborated by
\citet{2018MNRAS.475L..15G}, M22 \citep{2012Natur.490...71S} and M10 \citep{2018ApJ...855...55S}. 
In addition, our results match up closely with previous
hand-crafted models on the same simulation data. Among the 29 Galactic
GCs identified by hand in \citet{2018MNRAS.478.1844A}, 27 were also identified
by the machine learning classifier. We believe these results
signal that this technique is fairly accurate and that the
shortlisted clusters might be promising for observers to explore.

We have published our code publicly on Github for the sake of reproducibility
and reference. Installation and usage instructions are distributed along with
the code and can be found at \href{https://github.com/ammaraskar/black-holes-black-boxes}{github.com/ammaraskar/black-holes-black-boxes}. Additionally, the
repository links to an an interactive web page \revision{hosted using the binder project by }\cite{Binder} where one may put in their own
observational values to get a prediction from the classifier. As better observational data for Galactic and extraglactic GC will become available in the future, the classifier can be used with updated observational properties to determine whether the GCs could contain a sizeable number of BHs.

In the future, we plan to run a large survey of simulated GC models for the \textsc{mocca}-Survey Database II with improved stellar/binary evolution prescriptions and initial binary parameter distribution. This will comprise several thousand GC models that will sample a wider range of initial GC parameters compared to \textsc{mocca}-Survey Database I. With results from those simulations, the ML classifier can be retrained to better identify GCs that could contain a large population of BHs using their present-day properties. Furthermore, this would also open up more avenues for for ML applications, in particular a regression model could be trained to predict and approximate the number of compact objects in a cluster.

\section*{Acknowledgements}

We would like to thank the reviewer for an insightful report that helped improve the quality and readability of the paper. We would like to appreciate the work of Jawad Ahmad, Ikram Ullah Lali and Muhammad Latif who pointed out a mistake in Table A2 of the preprint while reproducing our results. We are very thankful that they communicated their findings to us. 
We are extremely grateful to Michela Mapelli for providing useful comments and suggestions that helped to improve the manuscript.
Many thanks are warranted for the scikit-learn community for creating such an 
extensive and useful open source project available for public consumption. 

AA(2) is supported by the Carl Tryggers Foundation through the grant CTS 17:113. MP and this project has received funding from the European Union's Horizon $2020$
research and innovation programme under the Marie Sk\l{}odowska-Curie grant agreement No. $664931$. MG acknowledge support from National Science Center (NCN), Poland, through the grant UMO-2016/23/B/ST9/02732. This work benefited from support by the International Space Science Institute (ISSI), Bern, Switzerland,  through its International Team programme ref. no. 393 {\it The Evolution of Rich Stellar Populations \& BH Binaries} (2017-18).




\bibliographystyle{mnras}
\bibliography{bibliography} 



\appendix

\section{Additional Tables And Figures}\label{appendix}

\begin{figure*}
	\centering
	\includegraphics[scale=0.375]{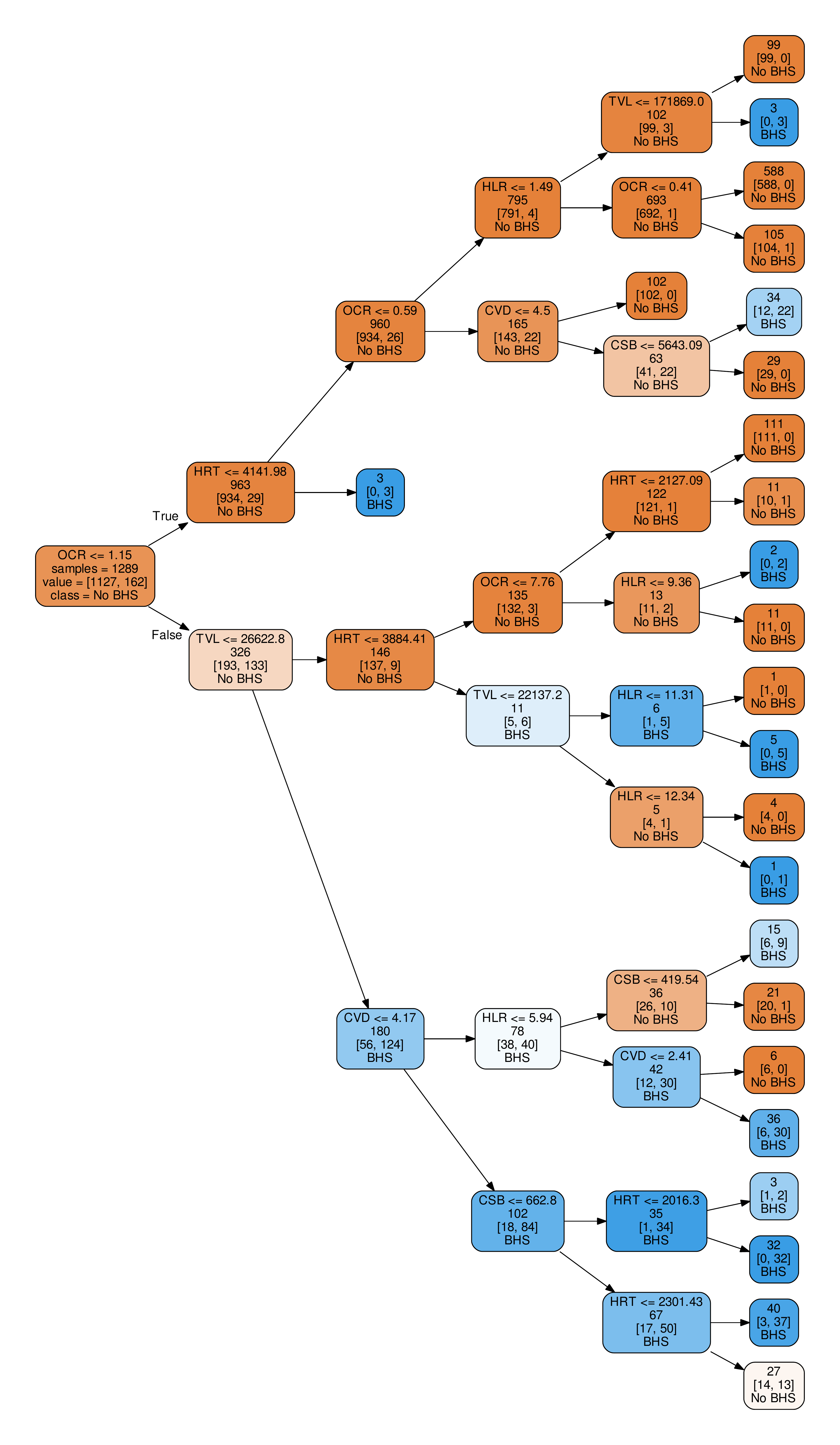}
    
    \caption{Visualization of decision tree. A legend is present at the root node. Nodes that are shaded blue are where the majority of examples contain a BHS. Nodes shaded orange are where the majority of example clusters do not contain a BHS.}
    
    \begin{tabular}{r@{: }l r@{: }l}
    HRT & Median Relaxation Time (Myr) & CVD & Central Velocity Dispersion (km s$^{-1}$) \\
    TVL & Total V-Band Luminosity (L$_{\odot}$) & CSB & Central Surface Brightness (m pc$^{-2}$)\\
    OCR & Observational Core Radius (pc) & OHLR & Half-Light Radius (pc)
    \end{tabular}
	\label{fig:decisiontree}
\end{figure*}

\begin{figure*}
	\centering
    
    \includegraphics[scale=1.0]{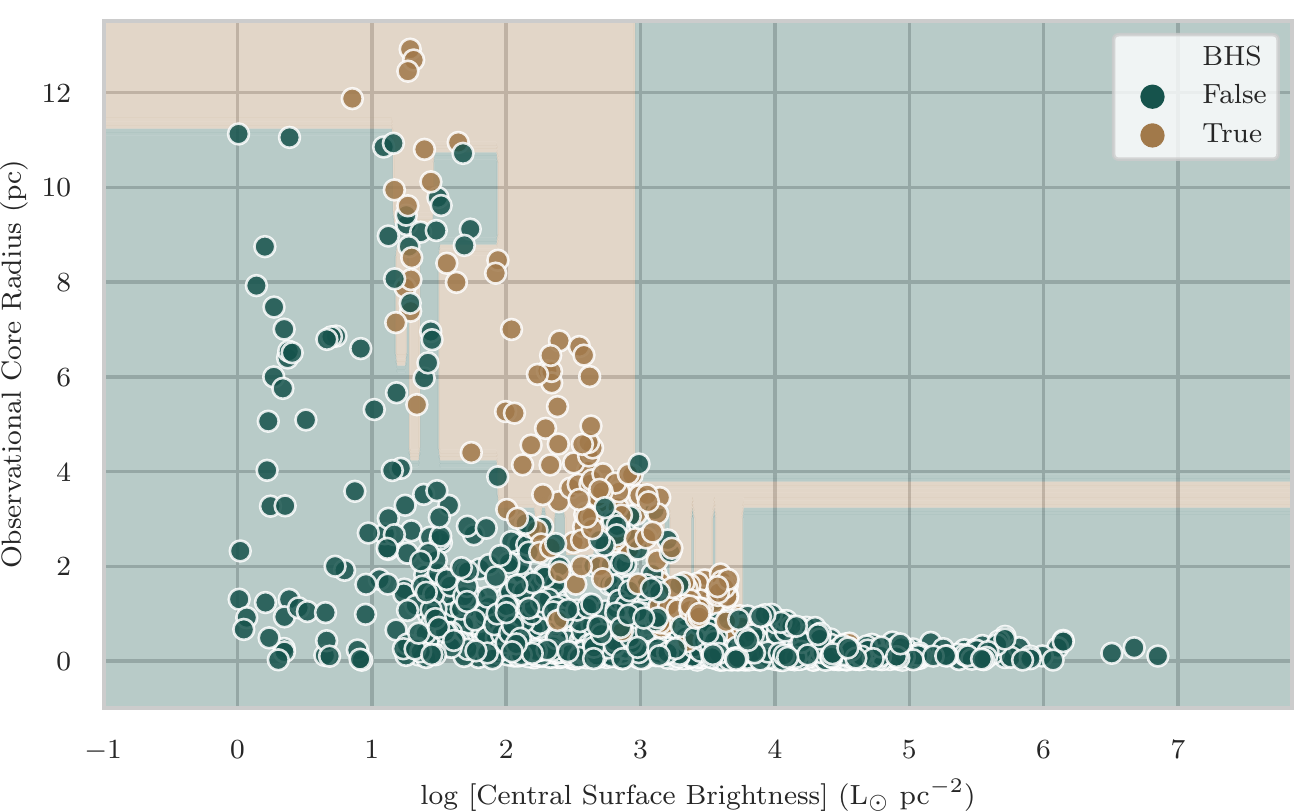}
    
    \caption{\revision{Decision boundaries (thresholds) for the two most important features in the decision tree classifier. The shading specifies where the classifier will predict that the cluster contains a BHS. Green shaded areas are where a prediction for no BHS will be given.}}
    \label{fig:decisionboundary}
\end{figure*}

The Appendix contains the full set of data and predictions along with larger figures 
related to the machine learning classifier. In particular, the following tables and
figures are presented:

\begin{description}
	\item[\textbf{\autoref{tab:harris}}] A full set of data and predictions from the \citet[][\emph{updated 2010}]{harris1996} catalogue.
    \item[\textbf{\autoref{tab:holger}}] A full set of data and predictions from the \cite{holger} catalogue.
    \item[\textbf{\autoref{fig:decisiontree}}] Visualization of the decision tree
    classifier as presented in \autoref{sec:decisiontrees}.
    \item[\textbf{\autoref{fig:decisionboundary}}] Visualization of decision boundaries
    within the decision tree classifier.
\end{description}

\newcommand{\tableHeader}{
Cluster & Half-Light & Central Surface & CVD \textsuperscript{$a$} & Total V-band & MRT \textsuperscript{$b$} & Core & BHS & BHS \\
    Name & Radius (pc) & Brightness ($\text{L}_{\odot}$ $\text{pc}^{-2}$) & (km $\text{s}^{-1}$) & Luminosity ($\text{L}_{\odot}$) & (Myr) & Radius (pc) & Prediction & (Fallback) \\
}

\begin{table*}
	\caption{Predictions for the presence of BH subsystems with data from the \citet[][\emph{updated 2010}]{harris1996} catalogue.}
    
    \centering
    
    \begin{tabular}{lllllllll}
    \toprule
    \tableHeader
    \midrule
	NGC 104 & 4.15 & 6.44 $\times 10^4$ & 11.00 & 5.01 $\times 10^5$ & 3548.13 & 0.47 & False & False\\
NGC 288 & 5.77 & 3.47 $\times 10^2$ & 2.90 & 4.29 $\times 10^4$ & 2089.30 & 3.50 & \textbf{True} & \textbf{True}\\
NGC 362 & 2.05 & 4.37 $\times 10^4$ & 6.40 & 2.01 $\times 10^5$ & 851.14 & 0.45 & False & False\\
NGC 1851 & 1.80 & 7.26 $\times 10^4$ & 10.40 & 1.84 $\times 10^5$ & 660.69 & 0.32 & False & False\\
NGC 2808 & 2.23 & 3.38 $\times 10^4$ & 13.40 & 4.88 $\times 10^5$ & 1412.54 & 0.70 & False & False\\
NGC 3201 & 4.42 & 9.14 $\times 10^2$ & 5.00 & 8.17 $\times 10^4$ & 1862.09 & 1.85 & \textbf{True} & \textbf{True}\\
NGC 4590 & 4.52 & 1.09 $\times 10^3$ & 2.50 & 7.59 $\times 10^4$ & 1862.09 & 1.74 & False & \textbf{True}\\
NGC 5139 & 7.56 & 6.87 $\times 10^3$ & 16.80 & 1.09 $\times 10^6$ & 12302.69 & 3.58 & \textbf{True} & \textbf{True}\\
NGC 5272 & 6.85 & 8.03 $\times 10^3$ & 5.50 & 3.05 $\times 10^5$ & 6165.95 & 1.10 & False & \textbf{True}\\
NGC 5286 & 2.48 & 1.18 $\times 10^4$ & 8.10 & 2.68 $\times 10^5$ & 1288.25 & 0.95 & False & \textbf{True}\\
NGC 5466 & 10.70 & 8.26 $\times 10^1$ & 1.70 & 5.30 $\times 10^4$ & 5754.40 & 6.66 & False & \textbf{True}\\
IC 4499 & 9.35 & 1.59 $\times 10^2$ & 2.50 & 7.24 $\times 10^4$ & 5370.32 & 4.59 & \textbf{True} & \textbf{True}\\
NGC 5904 & 3.86 & 1.33 $\times 10^4$ & 5.50 & 2.86 $\times 10^5$ & 2570.40 & 0.96 & False & \textbf{True}\\
NGC 5946 & 2.74 & 4.84 $\times 10^3$ & 4.00 & 6.37 $\times 10^4$ & 812.83 & 0.25 & False & False\\
NGC 6093 & 1.77 & 3.29 $\times 10^4$ & 12.40 & 1.67 $\times 10^5$ & 630.96 & 0.44 & False & False\\
NGC 6121 & 2.77 & 2.40 $\times 10^3$ & 4.00 & 6.43 $\times 10^4$ & 851.14 & 0.74 & False & False\\
NGC 6171 & 3.22 & 9.66 $\times 10^2$ & 4.10 & 6.03 $\times 10^4$ & 1000.00 & 1.04 & False & False\\
NGC 6205 & 3.49 & 8.41 $\times 10^3$ & 7.10 & 2.25 $\times 10^5$ & 1995.26 & 1.28 & False & \textbf{True}\\
NGC 6218 & 2.47 & 2.09 $\times 10^3$ & 4.50 & 7.18 $\times 10^4$ & 741.31 & 1.10 & \textbf{True} & \textbf{True}\\
NGC 6254 & 2.50 & 3.03 $\times 10^3$ & 6.60 & 8.39 $\times 10^4$ & 794.33 & 0.99 & \textbf{True} & \textbf{True}\\
NGC 6256 & 2.58 & 2.54 $\times 10^3$ & 6.60 & 6.19 $\times 10^4$ & 724.44 & 0.06 & False & False\\
NGC 6266 & 1.82 & 3.32 $\times 10^4$ & 14.30 & 4.02 $\times 10^5$ & 954.99 & 0.44 & False & False\\
NGC 6284 & 2.94 & 1.05 $\times 10^4$ & 6.30 & 1.31 $\times 10^5$ & 1230.27 & 0.31 & False & False\\
NGC 6293 & 2.46 & 1.87 $\times 10^4$ & 7.70 & 1.11 $\times 10^5$ & 870.96 & 0.14 & False & False\\
NGC 6341 & 2.46 & 2.36 $\times 10^4$ & 6.00 & 1.64 $\times 10^5$ & 1047.13 & 0.63 & False & False\\
NGC 6325 & 1.43 & 3.44 $\times 10^3$ & 5.90 & 5.20 $\times 10^4$ & 281.84 & 0.07 & False & False\\
NGC 6342 & 1.80 & 5.93 $\times 10^3$ & 5.20 & 3.16 $\times 10^4$ & 323.59 & 0.12 & False & False\\
NGC 6366 & 2.97 & 1.15 $\times 10^2$ & 1.30 & 1.69 $\times 10^4$ & 537.03 & 2.21 & False & False\\
NGC 6362 & 4.53 & 6.87 $\times 10^2$ & 2.80 & 5.15 $\times 10^4$ & 1584.89 & 2.50 & False & \textbf{True}\\
NGC 6388 & 1.50 & 5.82 $\times 10^4$ & 18.90 & 4.97 $\times 10^5$ & 794.33 & 0.35 & False & False\\
NGC 6397 & 1.94 & 2.36 $\times 10^4$ & 4.50 & 3.87 $\times 10^4$ & 398.11 & 0.03 & False & False\\
NGC 6441 & 1.92 & 4.18 $\times 10^4$ & 18.00 & 6.08 $\times 10^5$ & 1230.27 & 0.44 & False & False\\
NGC 6522 & 2.24 & 1.82 $\times 10^4$ & 6.70 & 9.82 $\times 10^4$ & 724.44 & 0.11 & False & False\\
NGC 6535 & 1.68 & 1.14 $\times 10^2$ & 2.40 & 6.79 $\times 10^3$ & 158.49 & 0.71 & False & False\\
NGC 6541 & 2.31 & 2.43 $\times 10^4$ & 8.20 & 2.19 $\times 10^5$ & 1071.52 & 0.39 & False & False\\
NGC 6558 & 4.63 & 9.31 $\times 10^3$ & 3.10 & 3.22 $\times 10^4$ & 1318.26 & 0.06 & False & False\\
NGC 6624 & 1.88 & 2.71 $\times 10^4$ & 5.40 & 8.47 $\times 10^4$ & 512.86 & 0.14 & False & False\\
NGC 6626 & 3.15 & 1.79 $\times 10^4$ & 8.60 & 1.57 $\times 10^5$ & 1479.11 & 0.38 & False & False\\
NGC 6656 & 3.13 & 3.92 $\times 10^3$ & 7.80 & 2.15 $\times 10^5$ & 1698.24 & 1.24 & \textbf{True} & \textbf{True}\\
NGC 6681 & 1.86 & 7.26 $\times 10^4$ & 5.20 & 6.03 $\times 10^4$ & 446.68 & 0.08 & False & False\\
NGC 6712 & 2.67 & 1.16 $\times 10^3$ & 4.30 & 8.55 $\times 10^4$ & 891.25 & 1.53 & False & \textbf{True}\\
NGC 6752 & 2.22 & 4.06 $\times 10^4$ & 4.90 & 1.06 $\times 10^5$ & 741.31 & 0.20 & False & False\\
NGC 6779 & 3.01 & 2.13 $\times 10^3$ & 4.00 & 7.87 $\times 10^4$ & 1023.29 & 1.20 & False & \textbf{True}\\
NGC 6809 & 4.45 & 6.56 $\times 10^2$ & 4.00 & 9.12 $\times 10^4$ & 1949.85 & 2.83 & False & \textbf{True}\\
NGC 6838 & 1.94 & 4.58 $\times 10^2$ & 2.30 & 1.50 $\times 10^4$ & 269.15 & 0.73 & False & False\\
NGC 6864 & 2.80 & 2.25 $\times 10^4$ & 10.30 & 2.29 $\times 10^5$ & 1412.54 & 0.55 & False & False\\
NGC 6934 & 3.13 & 4.18 $\times 10^3$ & 5.10 & 8.17 $\times 10^4$ & 1096.48 & 1.00 & \textbf{True} & \textbf{True}\\
NGC 7078 & 3.03 & 7.53 $\times 10^4$ & 13.50 & 4.06 $\times 10^5$ & 2089.30 & 0.42 & False & False\\
NGC 7089 & 3.55 & 1.77 $\times 10^4$ & 8.20 & 3.50 $\times 10^5$ & 2511.89 & 1.07 & False & \textbf{True}\\
NGC 7099 & 2.43 & 2.64 $\times 10^4$ & 5.50 & 8.17 $\times 10^4$ & 758.58 & 0.14 & False & False\\
    
	\midrule
    \multicolumn{9}{|c|}{\textit{(Less accurate models, those trained without Central Velocity Dispersion as a feature.)}} \\
    \midrule
1636-283 & 1.21 & 1.82 $\times 10^2$ & -- & 3.47 $\times 10^3$ & 74.13 & 1.21 & False & False\\
BH 176 & 4.95 & 1.65 $\times 10^1$ & -- & 3.60 $\times 10^3$ & 616.60 & 4.95 & False & False\\
E 3 & 4.95 & 2.09 $\times 10^1$ & -- & 3.80 $\times 10^3$ & 630.96 & 4.41 & False & False\\
HP 1 & 7.39 & 1.11 $\times 10^2$ & -- & 3.28 $\times 10^4$ & 2754.23 & 0.07 & False & False\\
    \multicolumn{9}{|c|}{\textbf{Continued on next page}} \\
    \midrule
    
    \end{tabular}
    
    {\footnotesize
    \textsuperscript{$a$}Central Velocity Dispersion \quad\textsuperscript{$b$}Median Relaxation Time
    }
             
	\label{tab:harris}
\end{table*}
\begin{table*}
	\centering
    
    \contcaption{}
    
    \begin{tabular}{lllllllll}
    \toprule
    \tableHeader
    \midrule
    IC 1276 & 3.74 & 8.33 $\times 10^1$ & -- & 3.98 $\times 10^4$ & 1071.52 & 1.59 & False & False\\
NGC 2298 & 3.08 & 1.00 $\times 10^3$ & -- & 2.86 $\times 10^4$ & 691.83 & 0.97 & False & False\\
NGC 4372 & 6.60 & 2.13 $\times 10^2$ & -- & 1.12 $\times 10^5$ & 3890.45 & 2.95 & \textbf{True} & \textbf{True}\\
NGC 4833 & 4.63 & 1.48 $\times 10^3$ & -- & 1.58 $\times 10^5$ & 2630.27 & 1.92 & False & \textbf{True}\\
NGC 5897 & 7.49 & 2.23 $\times 10^2$ & -- & 6.67 $\times 10^4$ & 3715.35 & 5.09 & \textbf{True} & \textbf{True}\\
NGC 5927 & 2.46 & 6.26 $\times 10^3$ & -- & 1.14 $\times 10^5$ & 870.96 & 0.94 & False & False\\
NGC 5986 & 2.96 & 3.11 $\times 10^3$ & -- & 2.03 $\times 10^5$ & 1513.56 & 1.42 & \textbf{True} & \textbf{True}\\
NGC 6101 & 4.70 & 2.79 $\times 10^2$ & -- & 5.11 $\times 10^4$ & 1659.59 & 4.35 & \textbf{True} & \textbf{True}\\
NGC 6139 & 2.50 & 4.80 $\times 10^3$ & -- & 1.89 $\times 10^5$ & 1122.02 & 0.44 & \textbf{True} & \textbf{True}\\
NGC 6144 & 4.22 & 3.47 $\times 10^2$ & -- & 4.70 $\times 10^4$ & 1380.38 & 2.43 & False & \textbf{True}\\
NGC 6235 & 3.35 & 8.18 $\times 10^2$ & -- & 2.81 $\times 10^4$ & 776.25 & 1.10 & False & False\\
NGC 6273 & 3.38 & 7.13 $\times 10^3$ & -- & 3.84 $\times 10^5$ & 2398.83 & 1.10 & False & \textbf{True}\\
NGC 6287 & 2.02 & 1.71 $\times 10^3$ & -- & 7.52 $\times 10^4$ & 562.34 & 0.79 & \textbf{True} & \textbf{True}\\
NGC 6304 & 2.44 & 4.80 $\times 10^3$ & -- & 7.11 $\times 10^4$ & 707.95 & 0.36 & False & \textbf{True}\\
NGC 6316 & 1.97 & 3.84 $\times 10^3$ & -- & 1.85 $\times 10^5$ & 776.25 & 0.51 & \textbf{True} & \textbf{True}\\
NGC 6333 & 2.21 & 4.03 $\times 10^3$ & -- & 1.29 $\times 10^5$ & 794.33 & 1.03 & \textbf{True} & \textbf{True}\\
NGC 6352 & 3.34 & 1.96 $\times 10^3$ & -- & 3.31 $\times 10^4$ & 831.76 & 1.35 & False & False\\
NGC 6355 & 2.36 & 3.20 $\times 10^3$ & -- & 1.45 $\times 10^5$ & 912.01 & 0.13 & False & False\\
NGC 6356 & 3.56 & 5.66 $\times 10^3$ & -- & 2.17 $\times 10^5$ & 1995.26 & 1.05 & False & \textbf{True}\\
NGC 6380 & 2.35 & 4.03 $\times 10^2$ & -- & 8.55 $\times 10^4$ & 724.44 & 1.08 & \textbf{True} & \textbf{True}\\
NGC 6401 & 5.89 & 1.20 $\times 10^3$ & -- & 1.24 $\times 10^5$ & 3388.44 & 0.77 & \textbf{True} & \textbf{True}\\
NGC 6402 & 3.52 & 1.57 $\times 10^3$ & -- & 3.73 $\times 10^5$ & 2454.71 & 2.14 & \textbf{True} & \textbf{True}\\
NGC 6426 & 5.51 & 3.35 $\times 10^2$ & -- & 3.98 $\times 10^4$ & 1905.46 & 1.56 & False & \textbf{True}\\
NGC 6440 & 1.19 & 5.46 $\times 10^3$ & -- & 2.70 $\times 10^5$ & 416.87 & 0.35 & \textbf{True} & \textbf{True}\\
NGC 6453 & 1.48 & 5.41 $\times 10^3$ & -- & 6.61 $\times 10^4$ & 331.13 & 0.17 & False & False\\
NGC 6496 & 3.35 & 3.11 $\times 10^2$ & -- & 6.49 $\times 10^4$ & 1096.48 & 3.12 & False & \textbf{True}\\
NGC 6517 & 1.54 & 3.44 $\times 10^3$ & -- & 1.71 $\times 10^5$ & 524.81 & 0.19 & False & \textbf{True}\\
NGC 6528 & 0.87 & 7.39 $\times 10^3$ & -- & 3.63 $\times 10^4$ & 114.82 & 0.30 & False & False\\
NGC 6539 & 3.86 & 1.35 $\times 10^3$ & -- & 1.77 $\times 10^5$ & 2089.30 & 0.86 & \textbf{True} & \textbf{True}\\
NGC 6544 & 1.06 & 1.09 $\times 10^4$ & -- & 5.11 $\times 10^4$ & 173.78 & 0.04 & False & False\\
NGC 6553 & 1.80 & 1.96 $\times 10^3$ & -- & 1.10 $\times 10^5$ & 524.81 & 0.93 & \textbf{True} & \textbf{True}\\
NGC 6569 & 2.54 & 2.00 $\times 10^3$ & -- & 1.75 $\times 10^5$ & 1122.02 & 1.11 & \textbf{True} & \textbf{True}\\
NGC 6584 & 2.87 & 3.23 $\times 10^3$ & -- & 1.02 $\times 10^5$ & 1047.13 & 1.02 & \textbf{True} & \textbf{True}\\
NGC 6637 & 2.15 & 6.81 $\times 10^3$ & -- & 9.73 $\times 10^4$ & 660.69 & 0.84 & False & False\\
NGC 6638 & 1.39 & 4.37 $\times 10^3$ & -- & 6.03 $\times 10^4$ & 288.40 & 0.60 & False & False\\
NGC 6642 & 1.72 & 6.81 $\times 10^3$ & -- & 3.94 $\times 10^4$ & 331.13 & 0.24 & False & False\\
NGC 6652 & 1.40 & 1.32 $\times 10^4$ & -- & 3.94 $\times 10^4$ & 245.47 & 0.29 & False & False\\
NGC 6717 & 1.40 & 7.13 $\times 10^3$ & -- & 1.57 $\times 10^4$ & 165.96 & 0.17 & False & False\\
NGC 6723 & 3.87 & 2.04 $\times 10^3$ & -- & 1.16 $\times 10^5$ & 1737.80 & 2.10 & \textbf{True} & \textbf{True}\\
NGC 6749 & 2.53 & 6.44 $\times 10^1$ & -- & 4.09 $\times 10^4$ & 602.56 & 1.42 & False & False\\
NGC 6760 & 2.73 & 1.25 $\times 10^3$ & -- & 1.17 $\times 10^5$ & 1023.29 & 0.73 & \textbf{True} & \textbf{True}\\
NGC 6981 & 4.60 & 8.97 $\times 10^2$ & -- & 5.60 $\times 10^4$ & 1698.24 & 2.27 & False & \textbf{True}\\
Pal 10 & 1.70 & 5.61 $\times 10^1$ & -- & 1.77 $\times 10^4$ & 234.42 & 1.39 & False & False\\
Pal 11 & 5.69 & 4.71 $\times 10^2$ & -- & 5.01 $\times 10^4$ & 2187.76 & 4.64 & \textbf{True} & \textbf{True}\\
Pal 12 & 9.51 & 7.53 $\times 10^2$ & -- & 5.25 $\times 10^3$ & 1905.46 & 0.11 & False & False\\
Pal 6 & 2.02 & 9.05 $\times 10^1$ & -- & 4.45 $\times 10^4$ & 436.52 & 1.11 & False & False\\
Pal 8 & 2.16 & 3.64 $\times 10^2$ & -- & 1.37 $\times 10^4$ & 295.12 & 2.09 & False & False\\
Terzan 1 & 7.44 & 3.35 $\times 10^0$ & -- & 4.97 $\times 10^3$ & 1318.26 & 0.08 & False & False\\
Terzan 10 & 2.62 & 1.12 $\times 10^2$ & -- & 2.96 $\times 10^4$ & 549.54 & 1.52 & False & False\\
Terzan 12 & 1.05 & 1.15 $\times 10^1$ & -- & 3.87 $\times 10^3$ & 60.26 & 1.16 & False & False\\
Terzan 2 & 3.32 & 4.98 $\times 10^1$ & -- & 1.92 $\times 10^4$ & 660.69 & 0.07 & False & False\\
Terzan 3 & 2.98 & 3.57 $\times 10^1$ & -- & 7.24 $\times 10^3$ & 371.54 & 2.81 & False & False\\
Terzan 5 & 1.45 & 1.35 $\times 10^2$ & -- & 7.94 $\times 10^4$ & 338.84 & 0.32 & False & \textbf{True}\\
Terzan 6 & 0.87 & 1.81 $\times 10^2$ & -- & 9.29 $\times 10^4$ & 165.96 & 0.10 & False & False\\
Terzan 7 & 5.11 & 1.98 $\times 10^2$ & -- & 8.63 $\times 10^3$ & 912.01 & 3.25 & False & False\\
Terzan 9 & 1.61 & 1.89 $\times 10^1$ & -- & 2.61 $\times 10^3$ & 100.00 & 0.06 & False & False\\
Ton 2 & 3.10 & 4.98 $\times 10^1$ & -- & 2.51 $\times 10^4$ & 660.69 & 1.29 & False & False\\
    \bottomrule
    \end{tabular}
    
\end{table*}

\renewcommand{\tableHeader}{
Cluster & Half-Light & Central Surface & CVD \textsuperscript{$a$} & Total V-band & HMRT \textsuperscript{$b$} & Core & BHS & BHS \\
    Name & Radius (pc) & Brightness ($\text{L}_{\odot}$ $\text{pc}^{-2}$) & (km $\text{s}^{-1}$) & Luminosity ($\text{L}_{\odot}$) & (Myr) & Radius (pc) & Prediction & (Fallback) \\
}

\begin{table*}
	\centering
	\caption{\revision{Predictions for the presence of BH subsystems with data from \citet{holger} catalogue for Milky Way GC parameters.}}
    
    \begin{tabular}{lllllllll}
    \toprule
    \tableHeader
    \midrule
    NGC 104 & 3.56 & 5.58 $\times 10^4$ & 13.78 & 4.40 $\times 10^5$ & 3388.44 & 0.49 & False & False\\
NGC 288 & 6.99 & 2.00 $\times 10^2$ & 3.72 & 4.90 $\times 10^4$ & 3801.89 & 4.78 & \textbf{True} & \textbf{True}\\
NGC 362 & 2.32 & 2.36 $\times 10^4$ & 8.36 & 2.09 $\times 10^5$ & 1995.26 & 0.47 & False & False\\
NGC 1851 & 1.65 & 1.57 $\times 10^5$ & 9.09 & 1.50 $\times 10^5$ & 1047.13 & 0.15 & False & False\\
NGC 2298 & 2.96 & 1.61 $\times 10^3$ & 1.46 & 2.52 $\times 10^4$ & 301.99 & 0.78 & False & False\\
NGC 2808 & 2.06 & 4.22 $\times 10^4$ & 14.05 & 4.52 $\times 10^5$ & 1621.81 & 0.68 & False & False\\
E 3 & 5.45 & 3.11 $\times 10^1$ & 1.56 & 3.79 $\times 10^3$ & 1819.70 & 2.77 & False & False\\
NGC 3201 & 3.77 & 9.93 $\times 10^2$ & 3.85 & 7.11 $\times 10^4$ & 2691.54 & 1.92 & \textbf{True} & \textbf{True}\\
NGC 4372 & 6.34 & 5.67 $\times 10^2$ & 5.00 & 1.32 $\times 10^5$ & 5888.44 & 4.82 & \textbf{True} & \textbf{True}\\
NGC 4590 & 4.51 & 7.85 $\times 10^2$ & 3.24 & 6.09 $\times 10^4$ & 3548.14 & 1.99 & False & False\\
NGC 4833 & 6.61 & 2.76 $\times 10^3$ & 4.82 & 2.96 $\times 10^5$ & 4168.69 & 2.36 & \textbf{True} & \textbf{True}\\
NGC 5139 & 7.04 & 5.72 $\times 10^3$ & 17.63 & 1.22 $\times 10^6$ & 24547.11 & 4.22 & False & \textbf{True}\\
NGC 5272 & 3.36 & 1.04 $\times 10^4$ & 7.07 & 2.53 $\times 10^5$ & 2951.21 & 1.04 & False & \textbf{True}\\
NGC 5286 & 2.64 & 1.82 $\times 10^4$ & 8.61 & 2.84 $\times 10^5$ & 1995.26 & 0.77 & False & \textbf{True}\\
IC 4499 & 9.83 & 1.81 $\times 10^2$ & 3.12 & 7.42 $\times 10^4$ & 11481.55 & 5.30 & \textbf{True} & \textbf{True}\\
NGC 5897 & 8.02 & 1.84 $\times 10^2$ & 3.09 & 6.66 $\times 10^4$ & 8317.64 & 5.85 & False & \textbf{True}\\
NGC 5904 & 3.61 & 8.51 $\times 10^3$ & 7.09 & 2.44 $\times 10^5$ & 2570.39 & 1.19 & False & \textbf{True}\\
NGC 5927 & 4.54 & 2.93 $\times 10^3$ & 5.78 & 1.35 $\times 10^5$ & 3019.95 & 1.45 & \textbf{True} & \textbf{True}\\
NGC 5986 & 2.44 & 5.77 $\times 10^3$ & 7.40 & 1.23 $\times 10^5$ & 1318.26 & 1.09 & False & \textbf{True}\\
NGC 6093 & 1.82 & 8.41 $\times 10^4$ & 9.92 & 1.74 $\times 10^5$ & 870.96 & 0.19 & False & False\\
NGC 6121 & 2.82 & 4.44 $\times 10^3$ & 4.65 & 5.64 $\times 10^4$ & 954.99 & 0.67 & False & False\\
NGC 6144 & 5.87 & 7.54 $\times 10^2$ & 1.55 & 8.57 $\times 10^4$ & 1737.80 & 2.72 & False & \textbf{True}\\
NGC 6139 & 2.62 & 3.10 $\times 10^4$ & 6.73 & 1.89 $\times 10^5$ & 1905.46 & 0.41 & False & False\\
Ter 3 & 5.27 & 7.14 $\times 10^1$ & 2.25 & 7.09 $\times 10^3$ & 2238.72 & 2.39 & False & False\\
NGC 6171 & 3.11 & 2.22 $\times 10^3$ & 3.61 & 4.02 $\times 10^4$ & 891.25 & 0.87 & False & False\\
NGC 6205 & 3.08 & 4.50 $\times 10^3$ & 8.09 & 1.74 $\times 10^5$ & 2290.87 & 1.58 & \textbf{True} & \textbf{True}\\
NGC 6218 & 2.81 & 5.49 $\times 10^3$ & 4.17 & 6.80 $\times 10^4$ & 812.83 & 0.71 & False & False\\
NGC 6254 & 2.78 & 5.55 $\times 10^3$ & 5.60 & 9.48 $\times 10^4$ & 1412.54 & 0.80 & False & \textbf{True}\\
NGC 6256 & 2.97 & 3.63 $\times 10^4$ & 3.96 & 3.23 $\times 10^4$ & 933.25 & 0.06 & False & False\\
NGC 6266 & 1.83 & 5.52 $\times 10^4$ & 16.27 & 2.76 $\times 10^5$ & 1348.96 & 0.38 & \textbf{True} & False\\
NGC 6273 & 3.13 & 1.29 $\times 10^4$ & 10.28 & 3.25 $\times 10^5$ & 3388.44 & 1.11 & False & \textbf{True}\\
NGC 6284 & 3.91 & 8.43 $\times 10^3$ & 6.08 & 1.61 $\times 10^5$ & 4677.35 & 0.63 & False & False\\
NGC 6293 & 2.76 & 6.42 $\times 10^4$ & 7.13 & 1.13 $\times 10^5$ & 1479.11 & 0.12 & False & False\\
NGC 6304 & 4.45 & 5.80 $\times 10^3$ & 5.52 & 2.02 $\times 10^5$ & 3090.29 & 1.02 & False & \textbf{True}\\
NGC 6316 & 2.72 & 1.61 $\times 10^4$ & 7.06 & 1.79 $\times 10^5$ & 2089.29 & 0.59 & False & False\\
NGC 6341 & 2.28 & 2.36 $\times 10^4$ & 7.60 & 1.48 $\times 10^5$ & 1737.80 & 0.61 & False & False\\
NGC 6342 & 2.12 & 3.99 $\times 10^4$ & 3.30 & 1.64 $\times 10^4$ & 354.81 & 0.06 & False & False\\
NGC 6356 & 3.71 & 8.14 $\times 10^3$ & 4.84 & 2.46 $\times 10^5$ & 3467.37 & 1.14 & False & \textbf{True}\\
NGC 6355 & 2.65 & 2.80 $\times 10^4$ & 3.74 & 1.26 $\times 10^5$ & 1230.27 & 0.33 & False & False\\
NGC 6352 & 3.24 & 2.84 $\times 10^3$ & 3.80 & 3.80 $\times 10^4$ & 1174.90 & 0.62 & False & False\\
NGC 6366 & 3.33 & 6.26 $\times 10^2$ & 3.29 & 2.02 $\times 10^4$ & 776.25 & 1.35 & False & False\\
HP 1 & 2.56 & 6.13 $\times 10^3$ & 5.08 & 2.20 $\times 10^4$ & 912.01 & 0.29 & False & False\\
NGC 6362 & 5.75 & 4.61 $\times 10^2$ & 4.19 & 5.67 $\times 10^4$ & 3311.31 & 2.87 & False & \textbf{True}\\
Lil 1 & 1.05 & 1.05 $\times 10^7$ & 20.03 & 4.52 $\times 10^5$ & 371.53 & 0.02 & False & False\\
Ton 2 & 4.46 & 4.62 $\times 10^2$ & 2.73 & 1.51 $\times 10^4$ & 1548.82 & 0.95 & False & False\\
NGC 6388 & 1.96 & 1.33 $\times 10^5$ & 19.06 & 5.49 $\times 10^5$ & 1905.46 & 0.35 & False & \textbf{True}\\
NGC 6402 & 3.57 & 5.46 $\times 10^3$ & 9.87 & 3.52 $\times 10^5$ & 3981.08 & 2.28 & \textbf{True} & \textbf{True}\\
NGC 6397 & 2.19 & 3.64 $\times 10^4$ & 5.23 & 4.08 $\times 10^4$ & 912.01 & 0.08 & False & False\\
Ter 5 & 1.12 & 3.26 $\times 10^5$ & 18.00 & 4.31 $\times 10^5$ & 363.08 & 0.19 & False & False\\
NGC 6440 & 1.25 & 1.96 $\times 10^5$ & 11.66 & 2.17 $\times 10^5$ & 416.87 & 0.15 & False & False\\
NGC 6441 & 2.03 & 1.09 $\times 10^5$ & 14.30 & 6.00 $\times 10^5$ & 2238.72 & 0.42 & False & False\\
Ter 6 & 1.49 & 8.23 $\times 10^5$ & 5.87 & 1.02 $\times 10^5$ & 288.40 & 0.05 & False & False\\
NGC 6496 & 5.05 & 2.98 $\times 10^2$ & 3.32 & 3.80 $\times 10^4$ & 2951.21 & 3.07 & False & False\\
NGC 6522 & 4.34 & 3.55 $\times 10^4$ & 6.09 & 1.18 $\times 10^5$ & 5623.41 & 0.15 & False & False\\
NGC 6535 & 2.08 & 2.43 $\times 10^3$ & 2.02 & 3.32 $\times 10^3$ & 218.78 & 0.14 & False & False\\
NGC 6528 & 1.97 & 3.52 $\times 10^4$ & 4.58 & 3.97 $\times 10^4$ & 467.74 & 0.12 & False & False\\
NGC 6539 & 4.04 & 6.30 $\times 10^3$ & 6.16 & 1.61 $\times 10^5$ & 3162.28 & 0.91 & False & False\\
NGC 6544 & 1.38 & 9.68 $\times 10^4$ & 3.09 & 3.37 $\times 10^4$ & 181.97 & 0.08 & False & False\\
NGC 6541 & 2.61 & 1.06 $\times 10^5$ & 5.71 & 1.94 $\times 10^5$ & 1513.56 & 0.21 & False & False\\
NGC 6553 & 1.94 & 9.77 $\times 10^3$ & 7.48 & 7.48 $\times 10^4$ & 660.69 & 0.56 & False & False\\
    \bottomrule
    \end{tabular}
    \label{tab:holger}
\end{table*}
\begin{table*}
	\centering
    \contcaption{}
    
    \begin{tabular}{lllllllll}
    \toprule
    \tableHeader
    \midrule
    NGC 6558 & 1.77 & 1.20 $\times 10^4$ & 2.74 & 2.78 $\times 10^4$ & 181.97 & 0.24 & False & False\\
IC 1276 & 3.28 & 1.29 $\times 10^3$ & 3.07 & 2.74 $\times 10^4$ & 954.99 & 0.92 & False & False\\
Ter 1 & 2.28 & 6.95 $\times 10^3$ & 7.10 & 6.93 $\times 10^4$ & 977.24 & 0.60 & False & False\\
NGC 6569 & 2.94 & 5.41 $\times 10^3$ & 5.76 & 1.53 $\times 10^5$ & 1949.84 & 1.21 & \textbf{True} & \textbf{True}\\
NGC 6624 & 1.49 & 4.28 $\times 10^4$ & 7.30 & 7.17 $\times 10^4$ & 251.19 & 0.19 & False & False\\
NGC 6626 & 1.76 & 3.71 $\times 10^5$ & 9.14 & 1.68 $\times 10^5$ & 776.25 & 0.13 & False & False\\
NGC 6642 & 1.50 & 5.12 $\times 10^4$ & 2.94 & 4.92 $\times 10^4$ & 234.42 & 0.13 & False & False\\
NGC 6656 & 3.23 & 6.08 $\times 10^3$ & 8.69 & 1.93 $\times 10^5$ & 2691.54 & 1.26 & False & \textbf{True}\\
NGC 6681 & 2.08 & 5.26 $\times 10^4$ & 8.18 & 5.68 $\times 10^4$ & 537.03 & 0.10 & False & False\\
NGC 6712 & 2.87 & 3.26 $\times 10^3$ & 4.76 & 9.00 $\times 10^4$ & 1047.13 & 1.23 & \textbf{True} & \textbf{True}\\
NGC 6715 & 3.20 & 8.52 $\times 10^4$ & 16.08 & 6.91 $\times 10^5$ & 5888.44 & 0.54 & False & False\\
NGC 6723 & 3.50 & 2.09 $\times 10^3$ & 4.36 & 8.85 $\times 10^4$ & 1348.96 & 1.68 & \textbf{True} & \textbf{True}\\
NGC 6749 & 3.38 & 1.20 $\times 10^3$ & 2.64 & 3.92 $\times 10^4$ & 954.99 & 1.28 & False & False\\
NGC 6752 & 2.40 & 4.30 $\times 10^4$ & 7.79 & 1.10 $\times 10^5$ & 1445.44 & 0.19 & False & False\\
NGC 6760 & 2.82 & 7.59 $\times 10^3$ & 5.55 & 1.11 $\times 10^5$ & 1737.80 & 0.75 & False & False\\
NGC 6779 & 4.38 & 3.17 $\times 10^3$ & 5.06 & 1.78 $\times 10^5$ & 3090.29 & 1.71 & \textbf{True} & \textbf{True}\\
NGC 6809 & 4.70 & 7.30 $\times 10^2$ & 3.91 & 7.90 $\times 10^4$ & 3235.94 & 2.93 & \textbf{True} & \textbf{True}\\
Pal 11 & 6.63 & 2.61 $\times 10^2$ & 3.21 & 6.55 $\times 10^4$ & 4365.16 & 5.08 & \textbf{True} & \textbf{True}\\
NGC 6864 & 2.46 & 2.25 $\times 10^4$ & 7.53 & 1.99 $\times 10^5$ & 2041.74 & 0.56 & False & False\\
NGC 6934 & 2.63 & 2.48 $\times 10^3$ & 3.42 & 6.65 $\times 10^4$ & 1479.11 & 1.12 & False & False\\
NGC 7078 & 1.90 & 2.45 $\times 10^0$ & 14.00 & 3.94 $\times 10^5$ & 1513.56 & 0.08 & \textbf{True} & \textbf{True}\\
NGC 7089 & 3.00 & 2.24 $\times 10^4$ & 12.35 & 3.59 $\times 10^5$ & 2754.23 & 0.77 & False & False\\
NGC 7099 & 2.44 & 6.65 $\times 10^4$ & 4.59 & 7.19 $\times 10^4$ & 1819.70 & 0.06 & False & False\\
Pal 12 & 6.98 & 2.05 $\times 10^1$ & 0.86 & 4.05 $\times 10^3$ & 1513.56 & 3.70 & False & False\\
    \bottomrule
    \end{tabular}
    
    {\footnotesize
    \textsuperscript{$a$}Central Velocity Dispersion \quad\textsuperscript{$b$}Half-Mass Relaxation Time
    }
\end{table*}

\clearpage

\bsp	
\label{lastpage}
\end{document}